\documentclass[]{relaxed_system_lab}

\usepackage[utf8]{inputenc}
\usepackage[T1]{fontenc}
\usepackage{geometry}
\usepackage{amsmath,amssymb}  %
\usepackage{charter}

\usepackage[toc,page,header]{appendix}

\usepackage{algorithm}
\usepackage{array}
\usepackage{wrapfig}

\usepackage{amsmath,amssymb} 
\DeclareMathOperator*{\argmin}{arg\,min}
\usepackage[utf8]{inputenc} 
\usepackage[T1]{fontenc}    
\usepackage{hyperref}       
\usepackage{url}            
\usepackage{booktabs}       
\usepackage{amsfonts}       
\usepackage[most]{tcolorbox}
\usepackage{nicefrac}       
\usepackage{microtype}      
\usepackage{xcolor} 
\usepackage{amssymb}
\usepackage{caption}
\usepackage{xspace}
\usepackage{multirow}
\usepackage{xcolor}
\usepackage{graphicx} 

\usepackage{xspace}
\usepackage{multirow}
\usepackage{seqsplit}

\usepackage{tcolorbox}
\usepackage{mdframed}
\mdfdefinestyle{insightbox}{
  backgroundcolor=green!5!white,
  linecolor=green!40!black,
  linewidth=0.3pt,
  roundcorner=2pt,
  innerleftmargin=4pt,
  innerrightmargin=4pt,
  innertopmargin=3pt,
  innerbottommargin=3pt
}
\mdfdefinestyle{motivationbox}{
  backgroundcolor=blue!5!white,
  linecolor=blue!40!black,
  linewidth=0.3pt,
  roundcorner=2pt,
  innerleftmargin=4pt,
  innerrightmargin=4pt,
  innertopmargin=3pt,
  innerbottommargin=3pt
}
\mdfdefinestyle{codebox}{
  backgroundcolor=black!4!white,
  linecolor=black!40!black,
  linewidth=0.3pt,
  roundcorner=2pt,
  innerleftmargin=4pt,
  innerrightmargin=4pt,
  innertopmargin=3pt,
  innerbottommargin=3pt
}

\usepackage[utf8]{inputenc} 
\usepackage[T1]{fontenc}    
\PassOptionsToPackage{hyphens}{url}
\usepackage{hyperref}       
\usepackage{xurl}           
\usepackage{booktabs}       
\usepackage{amsfonts}       
\usepackage{nicefrac}       
\usepackage{microtype}      
\usepackage{xcolor}         %
\usepackage{xspace}
\usepackage{amssymb}
\usepackage{tabularx}
\usepackage{caption}

\usepackage{amsmath}
\usepackage{algorithm}
\usepackage{algpseudocode}
\usepackage{graphicx}
\usepackage{subcaption}
\usepackage{wrapfig}

\ifdefined\final
\usepackage[disable]{todonotes}
\else
\usepackage[textsize=tiny]{todonotes}
\fi

\usepackage{amsmath,amssymb,bbm,mathtools}
\usepackage{graphicx,subcaption,booktabs,multirow}
\usepackage{xspace,microtype}
\usepackage{tikz}
\usepackage{hyperref}
\usepackage{algorithm}
\usepackage{wrapfig}
\usepackage{algpseudocode}
\usepackage{mdframed}
\mdfdefinestyle{insightbox}{
  backgroundcolor=green!5!white,
  linecolor=green!40!black,
  linewidth=0.3pt,
  roundcorner=2pt,
  innerleftmargin=4pt,
  innerrightmargin=4pt,
  innertopmargin=3pt,
  innerbottommargin=3pt
}
\usepackage{xspace}
\usepackage{seqsplit}
\usepackage{multirow}
\usepackage{seqsplit}
\newcommand{\sys}{\textsc{TurboServe}\xspace}
\newcommand{\companyname}{\text{Shengshu Technology}\xspace}
\newcommand{\jyh}[1]{{\color{black}#1}\xspace}
\usepackage{tcolorbox}
\usepackage{algorithm}
\usepackage{algpseudocode}
\usepackage{xcolor}
\usepackage{mdframed}
\usepackage{soul}
\mdfdefinestyle{insightbox}{
  backgroundcolor=green!5!white,
  linecolor=green!40!black,
  linewidth=0.3pt,
  roundcorner=2pt,
  innerleftmargin=4pt,
  innerrightmargin=4pt,
  nobreak=false,
  innertopmargin=3pt,
  innerbottommargin=3pt
}
\usepackage{wrapfig}
\mdfdefinestyle{motivationbox}{
  backgroundcolor=red!5!white,
  linecolor=red!40!black,
  linewidth=0.3pt,
  roundcorner=2pt,
  innerleftmargin=4pt,
  innerrightmargin=4pt,
  innertopmargin=3pt,
  innerbottommargin=3pt,
  nobreak=false,
}
\mdfdefinestyle{codebox}{
  backgroundcolor=black!4!white,
  linecolor=black!40!black,
  linewidth=0.3pt,
  roundcorner=2pt,
  innerleftmargin=4pt,
  innerrightmargin=4pt,
  innertopmargin=3pt,
  nobreak=false,
  innerbottommargin=3pt
}
\mdfdefinestyle{analysisbox}{
  backgroundcolor=blue!4!white,
  linecolor=blue!40!black,
  linewidth=0.3pt,
  roundcorner=2pt,
  innerleftmargin=4pt,
  innerrightmargin=4pt,
  innertopmargin=3pt,
  nobreak=false,
  innerbottommargin=3pt
}

\usepackage{natbib}
\usepackage{latexsym}

\usepackage{url}
\usepackage{amssymb}
\usepackage[utf8]{inputenc}
\usepackage{microtype}
\usepackage{booktabs}
\usepackage{pifont} 
\usepackage{multirow}
\usepackage{makecell}
\usepackage{paralist}
\usepackage{xspace}
\usepackage{color}
\usepackage{xcolor}
\usepackage{colortbl}
\usepackage{adjustbox}
\usepackage{hyperref} 
\usepackage[edges]{forest}
\usepackage{tikz} 
\usepackage{caption}
\usepackage{amsfonts}

\hypersetup{
    colorlinks,
    linkcolor={blue!80!black},
    citecolor={blue!80!black},
}
\tikzset{
    root/.style =             {align=center, text width=1cm, rounded corners=3pt, line width=0.3mm, fill=gray!10, draw=gray!80, font=\small},
    demographic/.style =         {align=center, text width=1.8cm, rounded corners=3pt, line width=0.3mm, fill=blue!10, draw=blue!80, font=\footnotesize},
    demographic_work/.style =    {align=center, text width=10cm, rounded corners=3pt, line width=0.3mm, fill=blue!10, draw=blue!0, font=\footnotesize},
    character/.style =         {align=center, text width=1.8cm, rounded corners=3pt, line width=0.3mm, fill=red!10, draw=red!80, font=\footnotesize},
    character_work/.style =    {align=center, text width=10cm, rounded corners=3pt, line width=0.3mm, fill=red!10, draw=red!0, font=\footnotesize},
    personalization/.style =           {align=center, text width=1.8cm, rounded corners=3pt, line width=0.3mm, fill=cyan!10, draw=cyan!80, font=\footnotesize},
    personalization_work/.style =      {align=center, text width=10cm, rounded corners=3pt, line width=0.3mm, fill=cyan!10, draw=cyan!0, font=\footnotesize},
    risk/.style =         {align=center, text width=1.8cm, rounded corners=3pt, line width=0.3mm, fill=orange!10, draw=orange!80, font=\footnotesize},
    risk_work/.style =    {align=center, text width=10cm, rounded corners=3pt, line width=0.3mm, fill=orange!10, draw=orange!0, font=\footnotesize},
}

\usepackage{CJK}

\newtcolorbox{promptbox}[1][]{
  enhanced,
  breakable,
  colback=promptboxlightgray,
  colframe=promptboxblue!30,
  arc=8pt,
  boxrule=0.5pt,
  left=12pt,
  right=12pt,
  top=8pt,
  bottom=8pt,
  fonttitle=\bfseries,
  fontupper=\linespread{1.2}\selectfont,
  title=#1
}

\title{\sys: Serving Streaming Video Generation Efficiently and Economically}

\author{Youhe Jiang$^1$$^2$$^*$, Haoxu Wang$^2$$^3$$^*$, Haotong Bao$^1$$^*$, Kai Jiang$^2$$^3$, Jianfei Chen$^2$$^3$, \\Jun Zhu$^2$$^3$, Fangcheng Fu$^1$, Jintao Zhang$^2$$^3$$^\dagger$}

\affiliation{$^1$Shanghai Jiao Tong University, $^2$Shengshu Technology, $^3$Tsinghua University}

\contribution{$^*$Equal contribution, $^\dagger$Corresponding author}

\abstract{
Streaming video generation is emerging as a new serving workload in which users interact with long-lived sessions that generate video progressively, chunk by chunk. Unlike offline video generation or typical LLM serving, streaming video generation must preserve session state across active and idle periods, repeatedly schedule ongoing sessions, and deliver each chunk under a tight latency target. This creates two key serving challenges in multi-user, multi-GPU environments: session duration heterogeneity, where long-running sessions make placement decisions \jyh{suboptimal} over time, and temporal user-demand heterogeneity, where \jyh{the number of active sessions} fluctuates sharply across bursts and idle periods. 

\vspace{0.5em}
We present \sys, the \textit{first} serving system designed specifically for streaming video generation workloads. \sys formulates serving as an online scheduling problem that jointly coordinates session placement and GPU provisioning. Its \textit{closed-loop scheduling algorithm} combines a migration-aware placement controller, which rebalances sessions across GPUs to reduce \jyh{the maximum} per-chunk latency, with a load-driven autoscaling controller, which adapts the GPU budget to workload variation for improved cost efficiency. To support these decisions at runtime, \sys implements coalesced chunk processing for batching concurrent active sessions on the same GPU, GPU-CPU offloading for session suspension and resumption, and NCCL-based GPU-GPU migration for online rebalancing.
We evaluate \sys on real-world production traces from \jyh{\companyname} across multiple model sizes and GPU clusters with up to 64 NVIDIA B300 GPUs. Compared with baseline serving configurations, \sys reduces worst-case per-chunk latency by 37.5\% and total GPU operating cost by 37.2\% on average. These results demonstrate the effectiveness of \sys in delivering cost-efficient, latency-stable serving for dynamic streaming video generation workloads. Our code is publicly available at \url{https://github.com/shengshu-ai/TurboServe}.
\vspace{-1.5em}
}

\begin{document}

\maketitle

\section{Introduction}
\label{sec:intro}

Video generation models, such as Sora~\cite{brooks2024video}, Veo~\cite{veo3_2025}, HunyuanVideo~\cite{kong2024hunyuanvideo}, Wan~\cite{wan2025wan}, CogVideoX~\cite{yang2025cogvideox}, and MovieGen \cite{polyak2024movie}, have achieved remarkable progress in synthesizing high-fidelity, temporally consistent videos from natural language prompts. These advances have unlocked a broad range of applications spanning film production, advertising, education, gaming, and personalized content creation, establishing video generation as one of the fastest-growing workloads in generative AI services~\cite{bruce2024genie,melnik2024video,valevski2025diffusion}.

Traditional video generation pipelines, such as Sora and HunyuanVideo, operate in an offline, one-shot fashion: the user submits a prompt and waits seconds to minutes for a complete \jyh{video} to be returned, with no opportunity to steer the output mid-generation. \jyh{This waiting time becomes increasingly impractical for long videos, where generation latency grows \jyh{quadratically} with video length and delays user feedback until the entire \jyh{video} is completed.} This \jyh{offline generation paradigm} is now being supplanted by a new wave of streaming video generation \jyh{pipelines}, such as StreamDiffusionV2~\cite{feng2025streamdiffusionv2}, Self-Forcing~\cite{huang2026self}, HYWorldPlay~\cite{hunyuanworld2025hy}, and LongLive~\cite{yang2025longlive}, among others~\cite{henschel2025streamingt2v,liang2025looking}, which generate video progressively chunk by chunk, accept user prompts and interaction signals on-the-fly, and emit each chunk under a tight per-chunk latency target so that users observe the video as it is being generated. \textit{This paradigm shift motivates a serving system purpose-built for streaming video generation.}

\begin{figure}
    \centering
    \includegraphics[width=0.5\linewidth]{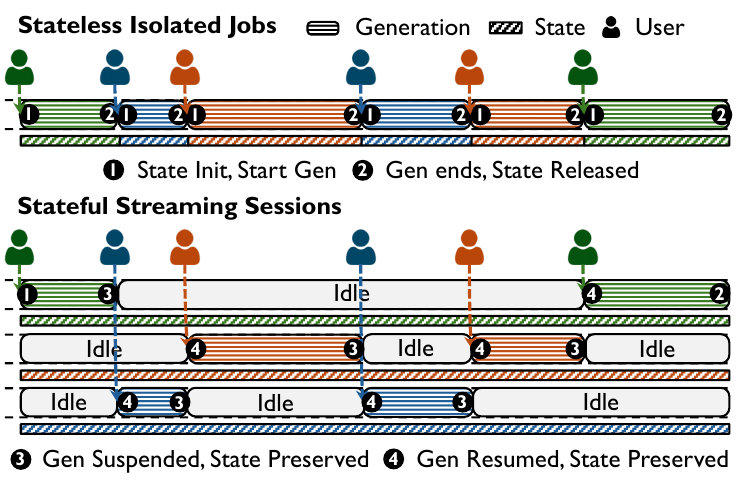}
    \caption{\jyh{Stateless isolated generation jobs (\textbf{\underline{top}}) compared with stateful streaming generation sessions (\textbf{\underline{bottom}}) in multi-user scenarios. In stateless generation, each user request initializes temporary generation state, starts generation, and releases the state once the generation job completes. In stateful streaming generation, each user corresponds to a persistent session that alternates between \emph{active generation} and \emph{idle} periods; generation can be suspended and later resumed while preserving the session state across periods.}}
    \label{fig:workflow}
\end{figure}

\begin{figure}[t!]
    \centering
    \includegraphics[width=0.5\linewidth]{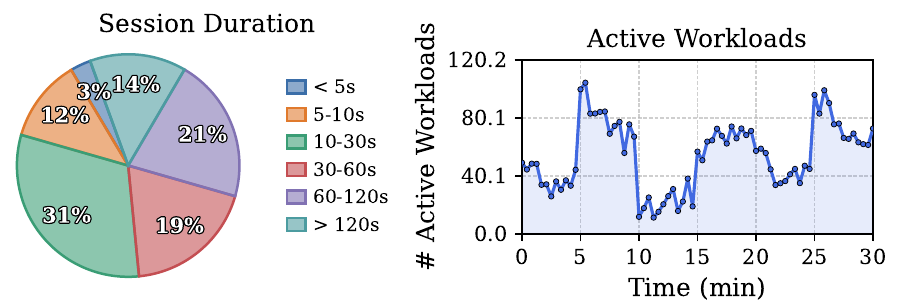}
    \caption{Example streaming generation workload characteristics from \companyname's production trace. \textbf{\underline{Left:}} distribution of session durations. \textbf{\underline{Right:}} number of active sessions over a 30-minute period.}
    \label{fig:workload-characteristics}
\end{figure}

Current diffusion-based video generation systems, such as \jyh{FastVideo~\cite{fastvideo}, xDiT~\cite{fang2024xdit}}, vLLM-Omni~\cite{yin2026vllm}, TurboDiffusion~\cite{zhang2025turbodiffusion}, and TridentServe \cite{xia2025tridentserve}, are primarily optimized for \jyh{stateless, one-shot generation requests}, where each request can be scheduled, resourced, and completed in isolation. Streaming video generation violates this \jyh{stateless serving assumption: serving is organized around persistent sessions that preserve prompt context and cached temporal states (typically, KV cache and related metadata), 
while continuously accepting new prompts or interaction signals. As a result, the system must coordinate evolving session states over time, rather than treating each generation as an isolated job, as demonstrated in~\autoref{fig:workflow}.} From this mismatch, building an efficient serving system for streaming video generation in a multi-user, multi-GPU environment introduces two primary dimensions of heterogeneity:
\begin{itemize}
\item \textbf{Challenge 1: Session duration heterogeneity.} Different sessions persist for vastly different durations, ranging from a single short clip to tens of minutes of iterative prompting and refinement, as illustrated in~\autoref{fig:workload-characteristics} (Left). 
Systems that treat sessions as fungible short-lived requests make placement decisions that become \jyh{suboptimal} as long-running sessions accumulate, causing some users to lose real-time generation capability.
\item \textbf{Challenge 2: Temporal user-demand heterogeneity.} Within and across sessions, demand fluctuates sharply over time: users alternate between bursts of activity and idle periods, and \jyh{the number of active sessions} rises and falls as new sessions arrive and existing ones complete, as illustrated in~\autoref{fig:workload-characteristics} (Right). \jyh{Systems with static provisioning face a fundamental trade-off: provisioning for peak demand over-provisions GPUs and wastes resources during low-demand periods, while provisioning for average demand under-provisions the system during bursts and prevents sessions from sustaining real-time generation.}
\end{itemize}


Resolving these challenges requires coordinated control over both session placement and GPU provisioning. On the session side, long-lived sessions need to be dynamically placed and migrated across GPUs so that load remains balanced as sessions arrive, depart, and transition between active and idle. On the resource side, the system needs to elastically adjust the provisioned GPU budget, scaling out during demand bursts to preserve latency headroom and scaling in during low-demand periods to avoid resource waste. Critically, these two controls are tightly coupled: the available GPU budget bounds which placements are feasible, while the current placement determines whether that budget is sufficient. Treating placement and autoscaling as independent components therefore leaves significant performance on the table, motivating a closed-loop design that jointly manages session placement and GPU provisioning for streaming video generation.

To address these challenges, we propose \sys, to the best of our knowledge, the \textit{first} serving system designed specifically for streaming video generation workloads in multi-user, multi-GPU environments. \sys jointly coordinates session placement and GPU provisioning in multi-session, multi-GPU environments with heterogeneous session durations and dynamic user demand. Our contributions are summarized as follows:

\begin{itemize}
    \item \textbf{\underline{Contribution 1:}} We formulate streaming video generation serving as an online scheduling problem that jointly determines session placement and GPU provisioning under per-chunk latency constraints. To solve this problem efficiently, we design a \textit{closed-loop scheduling algorithm} that combines a migration-aware placement controller with a load-driven autoscaling controller. Concretely, the placement controller performs event-driven min-max rebalancing, migrating sessions away from bottleneck GPUs to reduce the worst-case per-chunk latency; and the autoscaling controller adjusts the GPU budget based on runtime load feedback, provisioning additional GPUs during demand bursts and releasing GPUs during low-demand periods to improve cost efficiency.
    \item\textbf{\underline{Contribution 2:}} We implement \sys with runtime support for multi-session chunk execution and session mobility. \sys uses coalesced chunk processing to batch chunk generation for concurrent active sessions on the same GPU while preserving per-session states, and supports both GPU-CPU offloading for idle-session suspension/resumption as well as NCCL-based GPU-GPU migration for online rebalancing. These mechanisms improve GPU utilization, free GPU memory occupied by idle sessions, and enable fast load redistribution to reduce bottleneck per-chunk latency.
    \item\textbf{\underline{Contribution 3:}} We evaluate \sys on real-world production traces from \companyname across multiple video generation model sizes. We compare \sys against baseline serving configurations using two key metrics: worst-case per-chunk latency and total GPU operating cost. Experimental results show that \sys reduces worst-case per-chunk latency by 37.5\% and total GPU operating cost by 37.2\% on average, demonstrating its effectiveness in balancing serving latency and cost under dynamic streaming workloads.
\end{itemize}




\section{Background and Related Work}
\label{sec:background}

\noindent\textbf{Streaming video generation workflow.}
Streaming video generation serves a user session rather than a one-shot generation request (\autoref{fig:workflow}). In the workflow adopted by autoregressive streaming video models such as Self-Forcing~\cite{huang2026self}, the model generates video progressively: it first produces an initial video chunk from the user prompt, then generates subsequent chunks conditioned on the prompt together with the previously generated chunks and their cached state. During the session, the user may provide new prompts or interaction signals, or may become temporarily idle, causing the session to transition between active and idle periods. The serving system must therefore repeatedly schedule the same long-lived session, preserve its generation state, and deliver each video chunk within a tight latency target.

\vspace{0.25em}
\noindent\textbf{Streaming video generation.} Recent streaming diffusion and video-generation models have begun to shift video generation from offline batch synthesis toward online and interactive streaming. Systems such as StreamDiffusionV2~\cite{feng2025streamdiffusionv2}, StreamingT2V~\cite{henschel2025streamingt2v}, and StreamV2V~\cite{liang2025looking} demonstrate that diffusion-based video generation can be organized around continuous or autoregressive generation, enabling long videos, dynamic inputs, and interactive use cases. Concurrent efforts such as LongLive~\cite{yang2025longlive}, Self-Forcing~\cite{huang2026self}, and HYWorldPlay~\cite{hunyuanworld2025hy} further demonstrate this shift toward long-horizon, interactive, and real-time streaming video/world generation. These model-level advances create a new class of workloads in which video chunks must be generated continuously and delivered with low latency. However, most existing work focuses on model architecture, inference pipeline design, or frame-level generation quality, rather than multi-user serving efficiency at the cluster level. This leaves a gap between the capabilities of streaming video-generation models and the infrastructure needed to serve them efficiently at scale. \sys targets this gap by designing a system-level scheduling framework for streaming generation workloads.

\vspace{0.25em}
\noindent\textbf{LLM serving.} Recent LLM serving systems have significantly improved the efficiency of online inference~\cite{zheng2024sglang,jiang2023hexgen,zhang2025efficient,yu2022orca,zhong2024distserve,wu2024loongserve,mao2025skyserve,wu2023fast,ye2025flashinfer}. Representative systems such as vLLM~\cite{kwon2023efficient}, Sarathi-Serve~\cite{agrawal2024taming}, and AlpaServe~\cite{li2023alpaserve} consider each request as a finite
sequence of decoding iterations with dynamically growing
KV-cache state. \jyh{However, streaming video generation differs from typical LLM serving in two important aspects. First, \textit{state lifetime}: LLM serving usually maintains request state only until the response is completed, while streaming video generation must preserve session state across multiple active and idle periods. Second, \textit{latency target}: LLM serving commonly optimizes request completion time or token-level latency, while streaming video generation must satisfy a hard per-chunk latency target so that users can observe the video continuously as it is being synthesized. 
Existing LLM serving systems therefore do not explicitly optimize session-based streaming video generation workloads, where the serving system must jointly handle long-lived session state across active/idle periods and repeatedly schedule chunk generation under hard per-chunk latency targets.
}

\nocite{jiang2023hexgen,jiang2025hexgen,jiang2025demystifying,jiang2025thunderserve,zhang2025efficient}



\vspace{0.25em}
\noindent\textbf{Autoscaling for LLM serving.} Autoscaling and elastic resource management have become increasingly important for cost-efficient LLM serving~\cite{sun2024llumnix,xiang2025aegaeon,chen2025flashserve,li2025taming,fu2024serverlessllm}. Existing systems dynamically adjust GPU allocation, model placement, execution capacity, or checkpoint loading according to workload variation and latency objectives, substantially improving resource efficiency under fluctuating demand. However, these systems primarily target request-centric inference workloads, where requests are relatively short-lived and can often be scheduled independently. In contrast, streaming video generation introduces long-lived, stateful sessions that continuously alternate between active and idle periods while maintaining persistent generation states. As a result, scaling decisions affect not only the amount of available GPU resources, but also how long-lived session states are suspended, resumed, migrated, and redistributed during resource changes. This motivates a serving framework that jointly coordinates autoscaling and session scheduling under dynamic streaming workloads.

\nocite{tong2025parallax,peng2025hexgen,jiang2025cascadia,jiang2026boute,he2026efficient,jiang2026oserve,yao2026opentela,yan2025fsa}

\section{Motivation}
\label{sec:motivation}

Given the distinct aspects discussed in \S\ref{sec:background}, serving streaming video generation necessitates specific system optimizations.
In this section, we first introduce \textsc{TurboServe}\textsubscript{base}, which is the production runtime framework tailored for streaming video generation in \companyname.
Subsequently, we use a characterization study to motivate the design of \sys.



\subsection{\jyh{Basic Streaming Serving Runtime}}
\label{subsec:basic-runtime}

\jyh{This subsection introduces \textsc{TurboServe}\textsubscript{base}, which provides the basic runtime support required by streaming video generation: (\textbf{\underline{i}}) concurrent chunk execution under per-chunk latency constraints, (\textbf{\underline{ii}}) persistent session-state and lifecycle management, and (\textbf{\underline{iii}}) GPU-CPU state offloading for suspension and resumption.}

\jyh{
\vspace{0.25em}
\noindent\textbf{Concurrent chunk execution.}
Streaming video generation executes at the granularity of video chunks, where each GPU worker hosts a model replica and serves multiple active sessions assigned to that worker. At each chunk step, the runtime follows a coalesced execution procedure: (\textbf{\underline{i}}) it collects sessions whose next chunks are ready to be generated; (\textbf{\underline{ii}}) it groups ready sessions on the same GPU into a coalesced chunk batch; and (\textbf{\underline{iii}}) it invokes the model once for the batch and writes the generated chunks and updated states back to their corresponding sessions. Coalescing improves GPU utilization by amortizing model execution across concurrent sessions. However, as discussed in \S\ref{sec:background}, streaming serving requires each active session to deliver chunks within the target per-chunk latency, so the number of co-located sessions on each GPU must be bounded (detailed in~\S\ref{subsec:problemformulation}).
}

\jyh{
\vspace{0.25em}
\noindent\textbf{Session state and lifecycle.}
Unlike stateless one-shot generation jobs, streaming video generation maintains persistent session state across multiple active and idle periods. Accordingly, the runtime distinguishes three session states: (\textbf{\underline{i}}) \textit{execution}, where the session is assigned to a GPU and generates chunks; (\textbf{\underline{ii}}) \textit{suspend}, where the session is idle and its GPU slot is released while its state is preserved; and (\textbf{\underline{iii}}) \textit{terminate}, where the session has completed and all associated resources are released. These states are updated through lifecycle operations such as initialization, suspension, resumption, and termination, which are triggered by session arrivals, departures, and active/idle user transitions.
}

\jyh{
\vspace{0.25em}
\noindent\textbf{GPU-CPU state offloading.}
To avoid occupying GPU memory and execution slots for idle sessions, \textsc{TurboServe}\textsubscript{base} supports GPU-CPU state offloading for suspension and resumption.\footnote{We do not consider recomputation for state rematerialization, which is widely used in LLM serving. The primary reason is that video generation is generally compute-heavy while LLM serving is often bottlenecked by the memory-bound autoregressive generation.} The offloading procedure follows three steps: (\textbf{\underline{i}}) when a session becomes idle or must release its GPU slot, the runtime copies its persistent session state from GPU memory to host memory; (\textbf{\underline{ii}}) the session is marked as suspended and its GPU slot is released for other active sessions; and (\textbf{\underline{iii}}) when the session becomes active again, the state is restored to the selected GPU before chunk generation resumes. This mechanism decouples session lifetime from GPU residency: long-lived sessions can remain alive across idle periods without continuously occupying GPU resources, while preserving the context needed for later resumption.
}

\subsection{\jyh{Characterization and Motivation}}
\label{subsec:characterization}

This characterization focuses on understanding how session and GPU management affect the latency and cost of streaming video generation serving.

\begin{figure}[t!]
    \centering
    \includegraphics[width=0.5\linewidth]{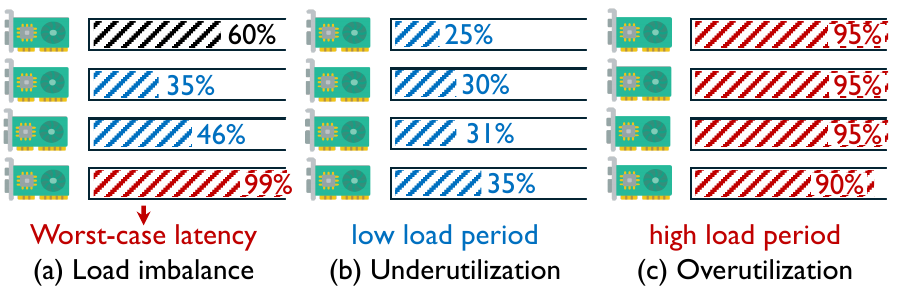}
    \caption{\jyh{Illustration of the observations. (\textbf{\underline{a}}) Evolving session activity causes load imbalance across GPUs, where the most heavily loaded GPU determines the worst-case per-chunk latency. (\textbf{\underline{b}}) During low-demand periods, a fixed GPU allocation leads to resource underutilization. (\textbf{\underline{c}}) During high-demand periods, the same static allocation becomes overutilized, resulting in high per-chunk latency.}}
    \label{fig:observation}
\end{figure}

\vspace{0.25em}
\noindent\textbf{Experiment setup.}
We evaluate on a cluster with 8 NVIDIA H100 GPUs. Each GPU can host multiple concurrent sessions, and each active session continuously generates video chunks under a target per-chunk latency constraint. We replay production traces from \companyname, which contain time-varying online arrivals, departures, and active/idle session transitions (detailed in~\autoref{tab:trace_stats} in~\autoref{appendix:workload-statistics}). We report two primary metrics: the worst-case per-chunk latency across active sessions and the total GPU operating cost over the trace. All configurations use the same model, trace, and latency target.


\vspace{0.25em}
\noindent\textbf{Baseline system configuration.} We first evaluate \jyh{\textsc{TurboServe}\textsubscript{base}} on the given experimental setup. In this baseline, each newly activated session is assigned to the GPU with the least current load among the available GPUs. Once assigned, the session remains on the same GPU until it is suspended or terminated. The system uses a fixed GPU allocation for the entire trace, providing a simple reference configuration for streaming video generation. This baseline reaches a worst-case per-chunk latency of \textbf{0.71 s}, with a total GPU cost of \textbf{3.99 \$}\footnote{\jyh{We run experiments on an in-house \companyname cluster, but report GPU operating cost in cloud-price-equivalent~\cite{aws_ec2_p5} dollars for readability.}}.

\vspace{0.5em}
\begin{mdframed}[style=motivationbox]
\small
\noindent\textbf{Observation: baseline configuration leads to latency imbalance and resource inefficiency.}
The baseline exposes two limitations. First, session activity evolves over time, so a placement that is initially balanced can become highly imbalanced as users transition between active and idle states. Since the worst-case per-chunk latency is determined by the most heavily loaded GPU, such imbalance directly increases system latency, as illustrated in~\autoref{fig:observation} (\textbf{\underline{a}}). Second, using a fixed GPU allocation throughout the trace leads to resource underutilization during low-demand periods while still incurring high latency during bursty periods, because the GPU budget cannot adapt to workload variation, as illustrated in~\autoref{fig:observation} (\textbf{\underline{b}}-\textbf{\underline{c}}).
\end{mdframed}
\vspace{0.25em}


\begin{figure}
    \centering
    \includegraphics[width=0.45\linewidth]{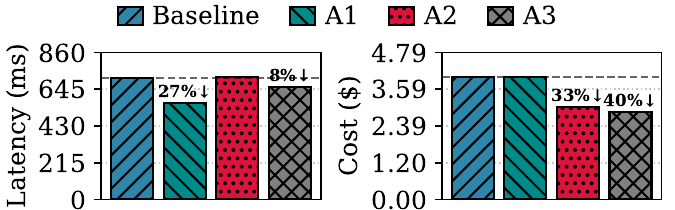}
    \caption{Case study on the impact of session and GPU management on latency and cost. \textit{Latency} is the worst-case per-chunk latency; \textit{Cost} is the total GPU operating cost. A1-3 represents Approach 1-3.}
    \label{fig:motivation_pic}
\end{figure}

\vspace{0.25em}
\noindent\textbf{Approach 1: session migration.}
We next evaluate a migration policy to isolate the benefit of session rebalancing. This policy uses the same fixed GPU allocation as the baseline, but periodically rebalances active sessions across GPUs every 10 seconds. At each rebalancing point, the system observes the current session distribution and migrates sessions to minimize the maximum per-GPU load. Since the GPU budget remains unchanged, any latency improvement comes from better session placement rather than additional resources. As shown in~\autoref{fig:motivation_pic}, compared with the baseline, session migration reduces the worst-case per-chunk latency by \textbf{26.53\%} while incurring the same GPU cost.

\vspace{0.5em}
\begin{mdframed}[style=insightbox]
\small
\noindent\textbf{\underline{Insight 1:} migration reduces bottleneck latency.}
Migration improves latency by moving sessions away from overloaded GPUs and equalizing load across the cluster. Periodic rebalancing prevents stale session placements from persisting after workload changes, especially when multiple sessions become active within a short interval.
\end{mdframed}
\vspace{0.25em}

\vspace{0.25em}
\noindent\textbf{Approach 2: GPU autoscaling.}
We next evaluate a GPU autoscaling policy to isolate the benefit of elastic GPU provisioning. Instead of maintaining a fixed GPU allocation throughout the workload trace, this policy dynamically adjusts the GPU budget according to the current workload demand. During high-demand periods, additional GPUs are provisioned to preserve latency headroom, while during low-demand periods unnecessary GPUs are released to reduce operating cost. As shown in~\autoref{fig:motivation_pic}, compared with the baseline, GPU autoscaling reduces the total GPU cost by \textbf{32.57\%} while maintaining the worst-case per-chunk latency within \textbf{0.71 s}.

\vspace{0.5em}
\begin{mdframed}[style=insightbox]
\small
\noindent\textbf{\underline{Insight 2:} autoscaling improves resource efficiency.}
Autoscaling improves cost efficiency by adapting the GPU budget to workload variation over time. Dynamic GPU provisioning preserves latency headroom during bursty periods while avoiding resource underutilization during low-demand periods.
\end{mdframed}
\vspace{0.25em}

\vspace{0.25em}
\noindent\textbf{Approach 3: joint migration and autoscaling.}
We finally evaluate a joint policy that combines session migration with GPU autoscaling. Similar to Approach 1, the policy periodically rebalances active sessions across GPUs every 10 seconds, while additionally performing rebalancing around each scaling decision according to the updated GPU set: scale-out adds GPUs and then redistributes active sessions to utilize the new capacity, while scale-in consolidates sessions before releasing unnecessary GPUs. As shown in~\autoref{fig:motivation_pic}, compared with the baseline, joint migration and autoscaling reduces the worst-case per-chunk latency by \textbf{8.17\%} and the total GPU cost by \textbf{40.25\%}.

\vspace{0.5em}
\begin{mdframed}[style=insightbox]
\small
\noindent\textbf{\underline{Insight 3:} migration and autoscaling should be coordinated.}
Migration and autoscaling address complementary bottlenecks: migration reduces load imbalance across GPUs, while autoscaling adapts the resource budget to workload demand. When combined, scale-out benefits from rebalancing to immediately use newly provisioned GPUs, and scale-in requires consolidation before GPUs can be safely released; meanwhile, balanced placement enables the load signal to better reflect true system demand rather than temporary imbalance introduced by stale placements, further improving the effectiveness of load-driven autoscaling.
\end{mdframed}
\vspace{0.25em}

\vspace{0.25em}
The three insights together motivate the design of \sys. Session migration reduces bottleneck per-chunk latency by balancing workload across GPUs, while GPU autoscaling improves resource efficiency under time-varying workload demand; a joint policy combines the benefits of both. Based on these observations, \sys integrates migration-aware placement and GPU autoscaling into a unified closed-loop scheduling framework for streaming video generation serving. We demonstrate the detailed system overview (\S\ref{sec:sched}), framework (\S\ref{sec:sched-framework}), and design (\S\ref{sec:systemdesign}) in the following sections.

\section{\sys Overview}
\label{sec:sched}

\begin{figure}
    \centering
    \includegraphics[width=0.6\linewidth]{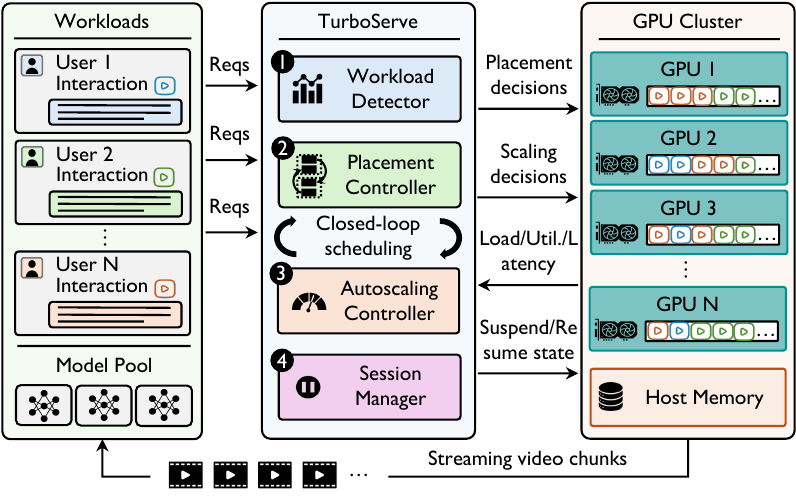}
    \caption{\sys system overview. \sys processes streaming requests and session events through a closed-loop scheduler that places and migrates sessions across GPU workers, adjusts GPU provisioning, and offloads suspended states to host memory. Runtime load, utilization, and latency feedback guide these decisions to balance serving latency and GPU cost.}
    \label{fig:system-overview}
\end{figure}

In this section, we demonstrate the system overview of \sys. The architecture overview of \sys is shown in~\autoref{fig:system-overview}. \sys consists of four major components: the workload detector, placement controller, autoscaling controller, and session manager.

\vspace{0.25em}
\noindent \textbf{Overall routine.} The overall routine of \sys is as follows. (\textbf{\underline{i}}) The \textit{workload detector} monitors recent session events, including arrivals, departures, and active/idle transitions, and extracts workload signals such as recent demand and demand variation using a sliding window. (\textbf{\underline{ii}}) The \textit{placement controller} uses the current session states, GPU status, and workload signals to assign sessions to GPU workers. It performs migration-aware rebalancing to reduce bottleneck latency across GPUs (detailed in~\S\ref{subsubsec:placementcontroller}). (\textbf{\underline{iii}}) The \textit{autoscaling controller} adjusts the GPU budget according to runtime feedback, including load, utilization, and per-chunk latency. It provisions or removes GPUs through scale-out and scale-in decisions (detailed in~\S\ref{subsubsec:autoscalingcontroller}). (\textbf{\underline{iv}}) The \textit{session manager} tracks session lifecycle states, including execution, suspension, termination, and migration (detailed in~\S\ref{subsec:session-management}). Active sessions are executed on GPUs to generate streaming video chunks, while suspended session states are stored in host memory and resumed when needed. By continuously feeding runtime measurements back to \sys, it jointly adapts session placement and GPU provisioning, thereby balancing per-chunk serving latency and GPU cost under dynamic streaming workloads.

\section{Scheduling Framework}
\label{sec:sched-framework}

\begin{table}[t]
\centering
\caption{Frequently used notations used in the scheduling framework.}
\small
\label{tab:notations}
\setlength{\tabcolsep}{6pt}
\renewcommand{\arraystretch}{1.05}
\resizebox{0.5\linewidth}{!}{
\begin{tabular}{l | p{6.5cm}}
\hline
\textbf{Notation} & \textbf{Description} \\
\hline

$\mathcal{S}(t)$ & Active session set at event $t$ \\
\hline
$\mathcal{G}(t)$ & Currently provisioned GPUs at event $t$ \\
\hline

$M(t)$ & Number of active GPUs at event $t$ \\
\hline
$\mathcal{C}(t)$ & GPU operating cost at event $t$ \\
\hline

$K$ & Max. concurrent sessions per GPU \\
\hline

$\phi_i(t)$ & Placement of session $s_i$ at event $t$ \\
\hline

$\alpha_i(t)$ & User-activity indicator for session $s_i$ at event $t$ \\
\hline

$\mathcal{L}(t)$ & Worst-case per-chunk serving latency \\
\hline

$\mathcal{L}^{*}(M,t)$ & Min. worst-case latency under budget $M(t)$ \\
\hline

$\rho_{\max}(t)$ & Max. normalized GPU load after placement \\
\hline

$\hat{\rho}(t)$ & Target per-GPU utilization \\
\hline

$\lambda(t)$ & Latency weight in the optimization objective \\
\hline

$M_{\mathrm{tar}}(t)$ & Target GPU budget computed by autoscaling \\

\hline
\end{tabular}
}
\end{table}

In this section, we first present the problem formulation for streaming-based serving in~\S\ref{subsec:problemformulation}. We then describe the placement controller for runtime session scheduling (\S\ref{subsubsec:placementcontroller}) and the autoscaling controller for GPU budget management (\S\ref{subsubsec:autoscalingcontroller}), which together constitute the closed-loop scheduling framework presented in~\S\ref{subsec:schedulingalgorithm}. \autoref{tab:notations} summarizes the essential notations used throughout this section.

\subsection{Problem Formulation}
\label{subsec:problemformulation}

We formulate the serving problem as an \textit{online scheduling problem} with two tightly coupled control components: (\textbf{\underline{i}}) session placement, which determines how active sessions are assigned to GPUs, and (\textbf{\underline{ii}}) cluster autoscaling, which controls the number of GPUs provisioned in the system. The system maintains a dynamically sized pool of GPUs, each capable of serving up to $K$ concurrent sessions without violating the per-chunk latency constraint. This induces a constrained resource allocation problem in which both the mapping of sessions to GPUs and the number of active GPUs must be continuously adjusted in response to workload dynamics. Our goal is to jointly minimize the cost of GPU provisioning and the per-chunk serving latency across all active sessions.

\vspace{0.25em}
\noindent\textbf{Event-driven scheduling.} We adopt an event-driven scheduling model to capture the online nature of the system. The scheduler is invoked upon system events, including session arrivals, departures, and state transitions (e.g., a user becoming active or idle). Each invocation corresponds to a decision epoch indexed by $t = 1, 2, \dots$, at which the scheduler observes the current system state (e.g., active sessions and resource utilization) and produces a scheduling decision. Between consecutive events, the system state evolves without intervention.

\vspace{0.25em}
\noindent\textbf{Sessions and states.} At each event $t$, sessions arrive and depart online; let $\mathcal{S}(t)$ denote the active session set with $|\mathcal{S}(t)| = N(t)$. Each session $s_i \in \mathcal{S}(t)$ is in one of three states: \textit{execution} (running on a GPU), \textit{suspend} (state offloaded to host memory, GPU slot released), or \textit{terminate} (session ended, all resources released and removed from the system). Let $\alpha_i(t) \in \{0, 1\}$ indicate whether the user is actively interacting with session $s_i$ at event $t$; transitions of $\alpha_i$ constitute state-change events that trigger the scheduler.

\vspace{0.25em}
\noindent\textbf{Decision variables.} At each event $t$, the scheduler jointly determines two quantities: (\textbf{\underline{i}}) the number of active GPUs $M(t) \in \{0, 1, \dots, M_{\max}\}$, which defines the set of active GPUs $\mathcal{G}(t) = \{g_1, \dots, g_{M(t)}\}$, and (\textbf{\underline{ii}}) a placement function $\phi_i(t)$ that assigns each session $s_i \in \mathcal{S}(t)$ to either an active GPU $g_j \in \mathcal{G}(t)$ (placing the session in the execution state) or to $\emptyset$ (placing it in the suspend state).

\vspace{0.25em}
\noindent\textbf{Optimization objective.} At each event $t$, the scheduler solves a \textit{multi-objective optimization problem}, aiming to jointly minimize the GPU operating cost and the per-chunk serving latency. Specifically, the operating cost is defined as $\mathcal{C}(t) = c_{\text{gpu}} \cdot M(t)$, where $c_{\text{gpu}}$ is the per-GPU rental cost, and $M(t)$ includes all provisioned GPUs, including those in the scale-out initialization phase (e.g., VM boot, model loading, and warm-up) that are not yet serving sessions. The latency is defined as the worst-case per-chunk latency across executing sessions, $\mathcal{L}(t) = \max_{s_i \in \mathcal{S}(t):\, \phi_i(t) \neq \emptyset} \ell_i(t)$, where $\ell_i(t)$ denotes the per-chunk latency of session $s_i$, which increases with per-GPU session co-location and may experience transient spikes during migration (e.g., due to state offloading, transfer, and resumption). The resulting optimization problem at event $t$ is:
\begin{equation}
\begin{aligned}
\label{eq:objective}
\argmin_{M(t),\; \phi(t)} \; & \mathcal{C}(t) + \lambda(t) \cdot \mathcal{L}(t), \\
\text{s.t.} \; &|\{i : \phi_i(t) = g_j\}| \leq K, && \forall g_j \in \mathcal{G}(t), \\
&\alpha_i(t) = 1 \implies \phi_i(t) \neq \emptyset, && \forall s_i \in \mathcal{S}(t).
\end{aligned}
\end{equation}
%
The first constraint enforces the per-GPU capacity limit, ensuring that no GPU serves more than $K$ sessions. The second constraint guarantees service responsiveness by requiring that any session receiving user input must be actively executed on a GPU.

\vspace{0.25em}
\noindent\textbf{Objective trade-off.} The weight $\lambda > 0$ balances the trade-off between cost and latency. Optimizing either objective in isolation leads to degenerate outcomes:
\begin{itemize}
    \item A large $\lambda$ reduces the problem to minimizing $\mathcal{L}(t)$, driving aggressive scale-out and migration that inflate $\mathcal{C}(t)$ without bound.
    \item A small $\lambda$ reduces the problem to minimizing $\mathcal{C}(t)$, driving $M(t)$ toward its minimum with sessions concentrated near $K$, which leaves no headroom for workload variation and easily pushes $\mathcal{L}(t)$ past the per-chunk latency deadline.
\end{itemize}
A properly tuned $\lambda$ avoids both failure modes, restraining $\mathcal{C}(t)$ while preserving sufficient slack in $\mathcal{L}(t)$ to absorb workload variation. The determination and adaptation of $\lambda$ are described in~\S\ref{subsubsec:autoscalingcontroller}.

\subsection{Closed-Loop Scheduling Algorithm}
\label{subsec:schedulingalgorithm}

We design a closed-loop scheduling algorithm that jointly optimizes session placement and cluster autoscaling. It consists of two tightly coupled components: (\textbf{\underline{i}}) a \textit{placement controller} that updates the assignment $\phi(t)$ and provides load feedback at each event $t$, and (\textbf{\underline{ii}}) an \textit{autoscaling controller} that adjusts the GPU budget $M(t)$ based on this feedback. Together, they form a closed control loop that regulates the system toward a desired operating point.

\subsubsection{Placement Controller}
\label{subsubsec:placementcontroller}

The placement controller operates at each event $t$ and determines the assignment $\phi(t)$ of sessions to the currently provisioned GPUs $\mathcal{G}(t)$. Given a fixed GPU budget $M(t)$, it approximately solves the placement optimization problem:
\begin{equation}
\mathcal{L}^{*}(M,t)=\argmin_{\phi(t)\ \text{feasible under }M(t)} \mathcal{L}(t),
\end{equation}
i.e., it seeks a feasible placement that minimizes the worst-case per-chunk latency under the current capacity while limiting unnecessary migration.

\vspace{0.25em}
\noindent\textbf{Initialization.} At each event $t$, the controller initializes from $\phi(t^{-})$ by removing terminated sessions and retaining existing assignments. This initialization enables incremental updates and limits unnecessary migration.

\vspace{0.25em}
\noindent\textbf{Session assignment.} The controller first handles sessions whose placement must be updated (e.g., newly arrived sessions and sessions that become active). The set of sessions requiring assignment is defined as:
\jyh{
\begin{equation}
U(t) =
\left\{
s_i \in \mathcal{S}(t) :
\alpha_i(t) = 1 \wedge \phi_i^{-}(t) = \emptyset
\right\}.
\label{eq:session-assignment-set}
\end{equation}
}
For each $s_i \in \mathcal{U}(t)$, the controller evaluates all feasible GPUs $g_j \in \mathcal{G}(t)$,
and selects the assignment that minimizes the resulting bottleneck latency $\mathcal{L}(t)$. When multiple assignments yield similar $\mathcal{L}(t)$, the controller selects among feasible GPUs using a fixed tie-breaking rule (e.g., preferring less-loaded GPUs).

\vspace{0.25em}
\noindent\textbf{Migration-aware min-max rebalancing.} After session assignment, the controller performs \textit{migration-aware min-max rebalancing} at each event $t$ to reduce the bottleneck latency $\mathcal{L}(t)$. Let $n_j(t) = |\{i : \phi_i(t)=g_j\}|$ denote the number of sessions assigned to GPU $g_j$, and let $g_{\max}(t) \in \arg\max_{g_j \in \mathcal{G}(t)} \hat{\ell}_j\big(n_j(t)\big)$ denote a GPU attaining the maximum per-chunk latency. The controller considers migrating a session $s_i$ currently assigned to $g_{\max}(t)$ to a target GPU $g_{j'} \in \mathcal{G}(t) \setminus \{g_{\max}(t)\}$, constructs the resulting placement $\phi'(t)$, and evaluates the new bottleneck latency $\mathcal{L}'(t)$. The migration cost $\kappa_i(t)$ captures the overhead of reassigning $s_i$ and is modeled using the $\alpha$-$\beta$ model~\cite{hockney2021computer}. The benefit of each candidate move is quantified by the gain function:
\begin{equation}
\Gamma_{i,j'}(t) = \mathcal{L}(t) - \mathcal{L}'(t) - \eta \cdot \kappa_i(t),
\end{equation}
where $\eta > 0$\footnote{Since migration overhead is minimized through system-level optimizations described in~\S\ref{sec:systemdesign}, $\eta$ is chosen to be small, enabling more aggressive rebalancing toward reducing bottleneck latency.} controls the trade-off between latency reduction and migration overhead. At each iteration, the controller selects the move $(i^*, j'^*) = \arg\max_{i,\, j'} \Gamma_{i,j'}(t)$ and applies it if $\Gamma_{i^*, j'^*}(t) > 0$, and repeats until no candidate move yields a positive gain. This event-driven local search progressively reduces the bottleneck latency while ensuring that migration is performed only when its benefit outweighs its cost.

\vspace{0.25em}
\noindent\textbf{Complexity analysis.} Let $N(t) = |\mathcal{S}(t)|$ and $M(t) = |\mathcal{G}(t)|$. The session assignment step evaluates up to $M(t)$ GPUs per session in $\mathcal{U}(t)$, yielding $O(|\mathcal{U}(t)| \cdot M(t))$ time. In rebalancing, each iteration considers at most $K$ sessions on the bottleneck GPU and evaluates their migration to up to $M(t)$ targets, resulting in $O(K \cdot M(t))$ time per iteration. The overall complexity is linear in $M(t)$ per event and negligible relative to the per-chunk generation latency.

\vspace{0.25em}
\noindent\textbf{Integration with autoscaling.} The placement controller is tightly coupled with the autoscaling controller, where rebalancing serves two distinct purposes:
\begin{itemize}
    \item \textit{Scale-in:} rebalancing consolidates sessions onto a reduced GPU set, ensuring that GPUs selected for removal are immediately freed, enabling prompt deprovisioning.
    \item \textit{Scale-out:} rebalancing redistributes sessions onto the expanded GPU set, ensuring that newly added GPUs are immediately utilized, enabling prompt load balancing.
\end{itemize}
We next describe the autoscaling controller in detail.

\begin{algorithm}[t]
\small
\caption{\small{Closed-Loop Scheduling Workflow}}
\label{alg:controller_interaction}
\begin{algorithmic}[1]
\Require Event stream $\{t\}$; session state $\mathcal{S}(t)$; placement $\phi(t)$; previous placement $\phi(t^{-})$; GPU budget $M(t)$; previous GPU budget $M(t^{-})$; load signal $\rho_{\max}(t)$; target GPU budget $M_{\mathrm{tar}}(t)$; placement controller $\textsc{Place}(\cdot)$; autoscaling controller $\textsc{Scale}(\cdot)$
\Ensure Updated placement $\phi(t)$; updated budget $M(t)$

\For{each event $t$}
    \Statex \hspace{\algorithmicindent} {\textcolor{green!80!blue}{/* Placement and GPU budget update */}}
    \State $(\phi(t), \rho_{\max}(t)) \gets \textsc{Place}(\mathcal{S}(t), \phi(t^{-}), M(t^{-}))$

    \State $M_{\mathrm{tar}}(t) \gets \textsc{Scale}(\rho_{\max}(t), M(t^{-}))$

    \If{$M_{\mathrm{tar}}(t) < M(t^{-})$}
        \Statex \hspace{\algorithmicindent} {\textcolor{green!80!blue}{/* Scale-in: rebalancing precedes removal */}}
        \State $(\phi(t), \rho_{\max}(t)) \gets \textsc{Place}(\mathcal{S}(t), \phi(t), M_{\mathrm{tar}}(t))$
        \State $M(t) \gets M_{\mathrm{tar}}(t)$
    \ElsIf{$M_{\mathrm{tar}}(t) > M(t^{-})$}
        \Statex \hspace{\algorithmicindent} {\textcolor{green!80!blue}{/* Scale-out: expansion precedes rebalancing */}}
        \State $M(t) \gets M_{\mathrm{tar}}(t)$
        \State $(\phi(t), \rho_{\max}(t)) \gets \textsc{Place}(\mathcal{S}(t), \phi(t), M(t))$
    \Else
        \Statex \hspace{\algorithmicindent} {\textcolor{green!80!blue}{/* No scaling: keep current budget */}}
        \State $M(t) \gets M(t^{-})$
    \EndIf
\EndFor
\end{algorithmic}
\end{algorithm}

\subsubsection{Autoscaling Controller}
\label{subsubsec:autoscalingcontroller}

The autoscaling controller determines the GPU budget $M(t)$ at each event $t$ based on feedback from the placement controller. Let $\rho_{\max}(t) = \max_{g_j \in \mathcal{G}(t)} n_j(t)/K$ denote the maximum normalized GPU load after placement, which reflects the current load pressure and remaining headroom under the current budget. While the placement controller approximately minimizes the worst-case per-chunk latency $\mathcal{L}^{*}(M,t)$ under a fixed GPU budget $M(t)$, the autoscaling controller adjusts $M(t)$ to minimize the overall objective:
\begingroup
\setlength{\abovedisplayskip}{1pt plus 0pt minus 0pt}
\setlength{\belowdisplayskip}{1pt plus 0pt minus 0pt}
\setlength{\abovedisplayshortskip}{0pt plus 0pt minus 0pt}
\setlength{\belowdisplayshortskip}{1pt plus 0pt minus 0pt}
\begin{equation}
\argmin_{M(t)} \; c_{\text{gpu}} \cdot M(t) + \lambda(t) \cdot \mathcal{L}^{*}(M,t),
\end{equation}
\vspace{-0.1em}
\endgroup
which is a decomposition of~\autoref{eq:objective}. Rather than solving this objective explicitly, the controller approximates it through target tracking by regulating the system toward a target utilization $\hat{\rho}(t)$, thereby balancing cost and latency through load-driven scaling decisions. Here, $\lambda(t)$ and $\hat{\rho}(t)$ jointly form the autoscaling controller's control parameters; their adaptive adjustment is described later in this section.

\vspace{0.25em}
\noindent \textbf{Hysteresis-based scaling trigger.} (\textit{When to scale?} )
Scaling decisions are driven by the load signal $\rho_{\max}(t)$ provided by the placement controller, relative to the target utilization $\hat{\rho}(t)$, which represents the desired fraction of the maximum capacity $K$ (e.g., $\hat{\rho}(t) = 0.7$ corresponds to 70\% utilization). To avoid oscillations under fluctuating workloads, the controller employs a \textit{hysteresis mechanism}~\cite{qu2018auto} with a tolerance parameter $\delta > 0$ (e.g., $\delta = 0.1$), which defines the stability margin around the target utilization: scale-out is triggered when $\rho_{\max}(t) > \hat{\rho}(t) + \delta$, and scale-in is triggered when $\rho_{\max}(t) < \hat{\rho}(t) - \delta$.

\vspace{0.25em}
\noindent\textbf{Proportional scaling adjustment.} (\textit{How much to scale?})
The scaling magnitude follows a \textit{proportional (target-tracking) policy}~\cite{kubernetes_hpa}, which adjusts the GPU budget in proportion to the deviation between the current load and the target utilization, enabling rapid convergence to the target utilization. Let $N_{\mathrm{req}}(t) = |\{i : \phi_i(t) \neq \emptyset\}|$ denote the total number of sessions requiring GPU execution. The target GPU budget is then computed as $M_{\mathrm{tar}}(t) = \left\lceil \frac{N_{\mathrm{req}}(t)}{K \hat{\rho}(t)} \right\rceil$, and the scaling adjustment is $\Delta M(t) = M_{\mathrm{tar}}(t) - M(t)$. After each scaling action, the updated load signal $\rho_{\max}(t)$ is obtained from the placement controller after recomputing $\phi(t)$ under the new budget, and is then used for subsequent decisions.

\vspace{0.25em}
\noindent\textbf{Adaptive control parameters.} (\textit{How to adapt?})
The autoscaling controller regulates the trade-off between GPU cost and worst-case per-chunk latency through the joint adjustment of the control parameters $(\lambda(t), \hat{\rho}(t))$, where $\lambda(t)$ controls the latency weight in~\autoref{eq:objective} and $\hat{\rho}(t)$ specifies the target per-GPU utilization. To adapt to workload dynamics, the controller dynamically adjusts $(\lambda(t), \hat{\rho}(t))$ according to recent workload characteristics (e.g., session arrivals and activation patterns)\footnote{Following the workload-classification paradigm of Quasar~\cite{delimitrou2014quasar}, we classify workloads into a small set of levels based on a sliding-window volatility metric over session activations, profile offline the cost-optimal control parameters for each level, and at runtime apply the parameters associated with the level of the current workload. The full procedure and a comparison against an offline trace-tuned oracle (near-optimality validation) are detailed in~\autoref{app:volatility-mapping}.}:
\begin{itemize}
    \item During highly fluctuating periods, the controller selects a larger $\lambda(t)$ and a smaller $\hat{\rho}(t)$ to preserve additional headroom and avoid latency spikes.
    \item During stable periods, the controller selects a smaller $\lambda(t)$ and a larger $\hat{\rho}(t)$ to improve resource efficiency.
\end{itemize}
As a result, the control parameters $(\lambda(t), \hat{\rho}(t))$ evolve over time, enabling the autoscaling controller to adapt the system between latency-oriented and cost-oriented operating regimes in response to workload dynamics.

\begin{figure}
    \centering
    \includegraphics[width=0.5\linewidth]{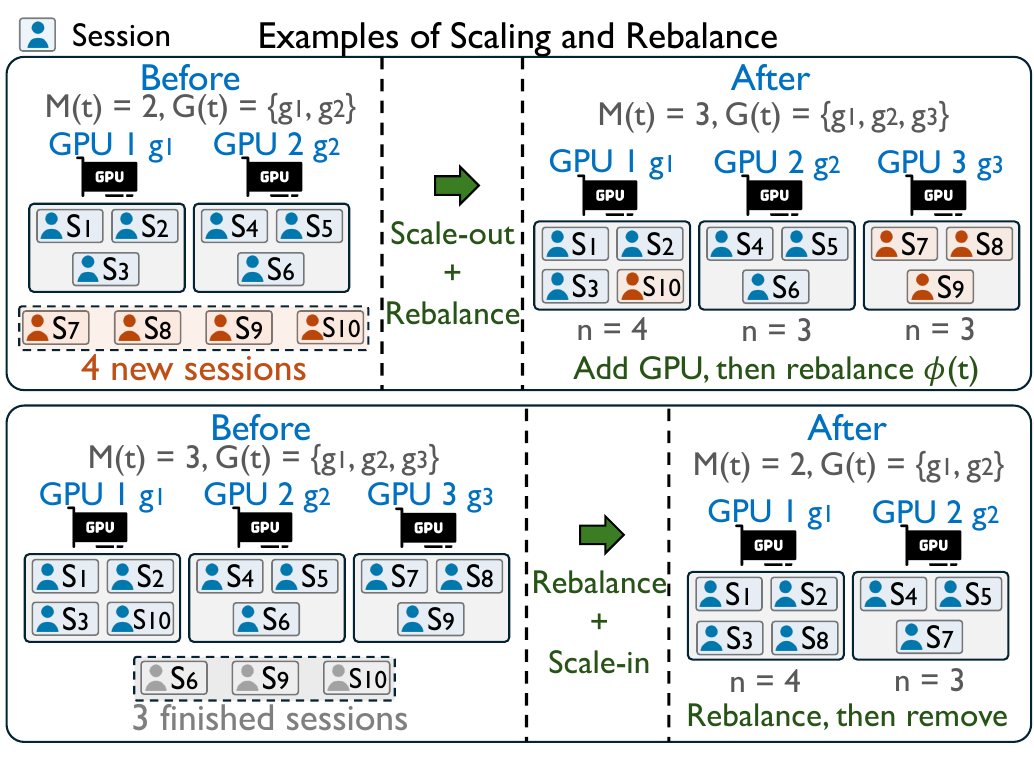}
    \caption{Illustrative examples of closed-loop GPU autoscaling and session rebalancing. \jyh{The top example shows scale-out followed by session rebalancing, where new GPUs are added and sessions are redistributed to reduce load concentration. The bottom example shows rebalancing followed by scale-in, where sessions are first consolidated onto fewer GPUs before underutilized GPUs are removed.}}
    \label{fig:session}
\end{figure}

\vspace{0.25em}
\noindent\textbf{Put it together.} The interaction between the placement controller and autoscaling controller constitutes a closed control loop (as demonstrated in~\autoref{alg:controller_interaction} and illustrated in~\autoref{fig:session}). In this loop, the placement controller provides the load feedback $\rho_{\max}(t)$, which informs autoscaling decisions, while the autoscaling controller updates the GPU budget $M(t)$, thereby regulating system load and worst-case per-chunk latency under dynamic workloads.

\section{\jyh{\sys Deployment}}
\label{sec:systemdesign}

\jyh{Building on the basic streaming serving runtime introduced in~\S\ref{subsec:basic-runtime}, this section describes the deployment-level mechanisms that make \sys efficient under dynamic multi-session workloads. Since the scheduling framework in~\S\ref{sec:sched-framework} continuously updates session placement and GPU provisioning, the runtime must efficiently manage session states and cluster resources during these updates. \sys therefore focuses on two implementation aspects: \textit{session management} and \textit{GPU management}.}

\subsection{\jyh{Session Management}}
\label{subsec:session-management}

\jyh{
\noindent\textbf{Session memory layout.}
Each GPU worker organizes runtime memory into three parts: (\textbf{\underline{i}}) a shared model replica used by all sessions on the worker; (\textbf{\underline{ii}}) isolated per-session state regions that store the session descriptor, prompt/control embeddings, temporal or KV-cache states, chunk-history features, intermediate latent buffers, and output metadata; and (\textbf{\underline{iii}}) a session ownership table that maps each active session to its current worker and state buffers. This layout separates static model resources from dynamic session state. As a result, \sys can migrate or offload a session by moving only its per-session state region, without moving the model replica or completed output chunks.
}

\jyh{
\vspace{0.25em}
\noindent\textbf{GPU-GPU state migration.}
When the placement controller decides to move an active session from one GPU to another, \sys migrates only the corresponding per-session state region. To support low-overhead transfer, each worker registers the memory space used for session states with the communication runtime. The target worker can then fetch the registered state buffers from the source worker through RDMA/NIXL-style one-sided GPU memory access~\cite{gpudirect_rdma,daoud2016gpurdma,nixl} and install them into its local session table. This avoids unnecessary CPU staging and keeps migration scoped to the state needed to continue generation.
}

\vspace{0.25em}
\noindent\textbf{Migration consistency.}
\sys performs GPU-GPU migration only at chunk boundaries to avoid interrupting an in-flight generation step. The consistency protocol follows three steps: (\textbf{\underline{i}}) the source worker completes the current chunk update and freezes the session state to be migrated; (\textbf{\underline{ii}}) the target worker fetches the state and verifies that the required buffers have been installed; and (\textbf{\underline{iii}}) session ownership is updated only after the transfer completes, ensuring that future chunks are generated on the target GPU and preventing duplicated execution. This design preserves session correctness while enabling fast rebalancing of active sessions to reduce bottleneck per-chunk latency.

\subsection{GPU Management}
\label{subsec:resource-management}

\jyh{
\noindent\textbf{\companyname-managed GPU pool.}
\sys is deployed on top of \companyname's internal GPU cluster management system. The cluster manager maintains a pool of available GPU workers whose container images, runtime dependencies, and model checkpoints are pre-staged on local storage. \sys does not directly manage physical machines; instead, it maintains a logical serving pool consisting of GPU workers currently admitted for streaming video generation. Each active worker hosts a model replica and periodically reports its load, health status, and availability to the autoscaling controller (\S\ref{subsubsec:autoscalingcontroller}). This separation allows \sys to make serving-level scaling decisions while relying on \companyname's existing infrastructure for resource allocation, isolation, and cleanup.
}

\jyh{
\vspace{0.25em}
\noindent\textbf{Scale-out procedure.}
When the autoscaling controller triggers scale-out, \sys requests additional GPU workers from the \companyname cluster manager. The scale-out procedure follows two steps: (\textbf{\underline{i}}) the cluster manager reserves available GPUs and attaches them to the \sys serving pool; and (\textbf{\underline{ii}}) \sys launches the worker runtime, loads the model replica from the locally staged checkpoint, and marks the worker ready for session assignment. Once a worker is ready, the placement controller (\S\ref{subsubsec:placementcontroller}) can assign new sessions or migrate existing sessions to the added GPU. Since images and checkpoints are already staged by the cluster manager, scale-out avoids expensive remote image pulling or checkpoint downloading on the critical path.
}

\jyh{
\vspace{0.25em}
\noindent\textbf{Scale-in procedure.}
When the autoscaling controller triggers scale-in, \sys marks selected workers as draining before releasing them. Draining workers no longer accept new sessions, and the placement controller migrates active sessions to remaining GPUs or offloads idle sessions through the session-state migration mechanism in \S\ref{subsec:session-management}. After all resident sessions have been moved or suspended, \sys unloads the model replica, frees GPU memory, and removes the worker from the serving pool. The released GPU is then returned to the \companyname cluster manager and becomes available for other workloads.
}

\vspace{-0.5em}
\section{System Evaluation}
\label{sec:eval}

\subsection{Experimental Setup}
\label{subsec:setup}
\textbf{Hardware environments.} We evaluate \sys on \companyname’s internal GPU serving clusters consisting of 16 NVIDIA H20 GPUs (\textbf{\underline{Cluster 1}}) and 64 NVIDIA B300 GPUs (\textbf{\underline{Cluster 2}}) for streaming video generation deployment. Each server contains 8 GPUs connected through high-bandwidth NVLink interconnects (956 GB/s), while cross-node communication is supported through RDMA-enabled InfiniBand networking (50 GB/s).

\vspace{0.25em}
\noindent\textbf{Models and workloads.} We evaluate \sys using LongLive-style streaming video generation models with multiple model sizes (LongLive-1.3B and its larger variants). The workload is replayed from \companyname’s production traces, which contain heterogeneous session durations, bursty online activation patterns, and time-varying numbers of active sessions. Detailed statistics of the evaluated traces (Trace 1-6) are provided in~\autoref{appendix:workload-statistics}.

\begin{figure*}[t!]
    \centering
    \includegraphics[width=\linewidth]{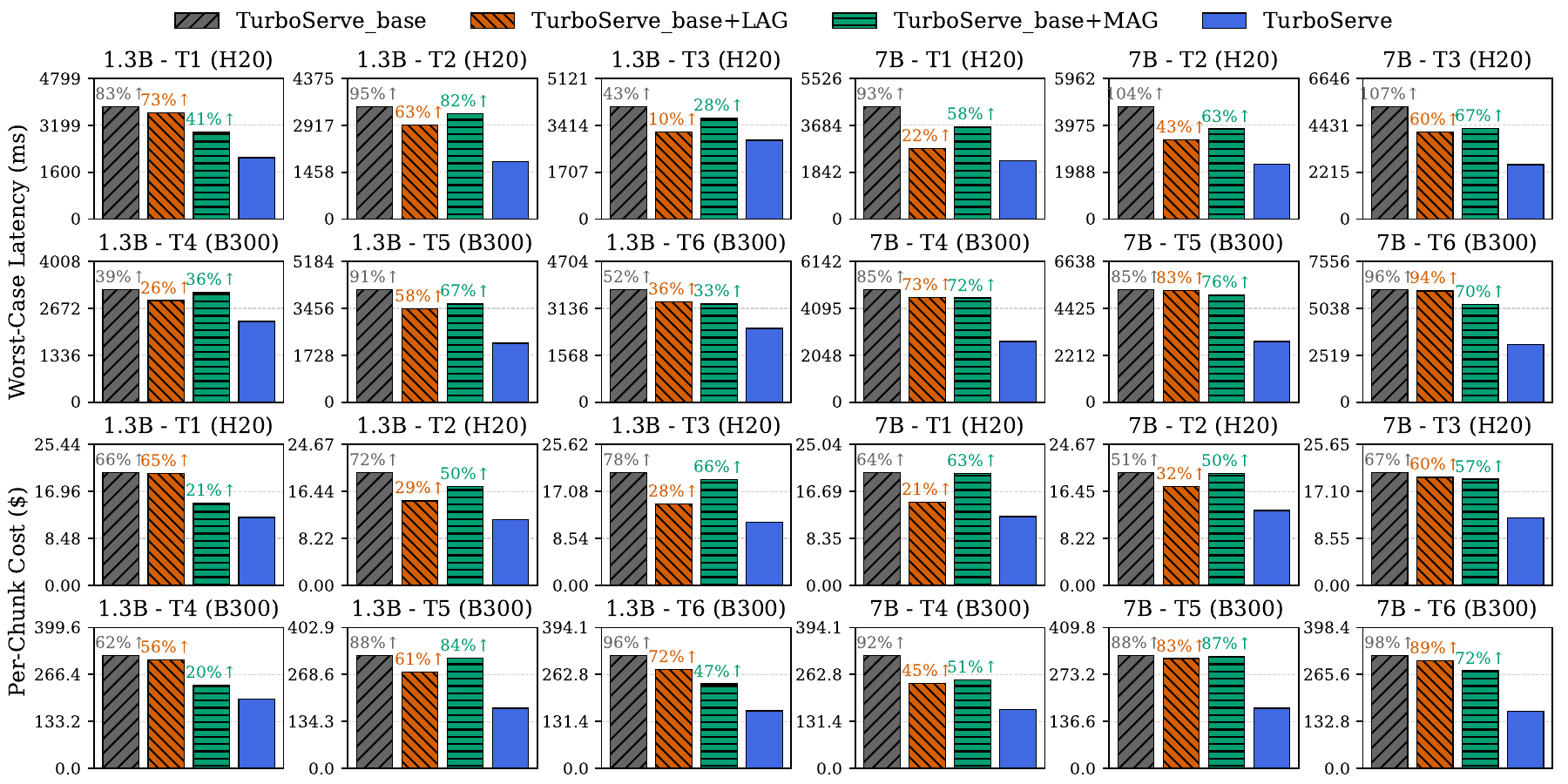}
    \caption{End-to-end experimental results comparing \sys against baseline systems across different models, traces, and clusters. The first two rows report the maximum per-chunk latency under the same cost budget, while the last two rows report the per-chunk cost under the same latency constraint.}
    \label{fig:e2e}
\end{figure*}

\vspace{0.25em}
\noindent\textbf{Evaluation metrics.} We evaluate \sys using two primary metrics: (\textbf{\underline{i}}) worst-case per-chunk latency, defined as the maximum chunk-generation latency across all active sessions, which captures the latency bottleneck perceived by users; and (\textbf{\underline{ii}}) GPU operating cost, defined as the total accumulated cost of provisioned GPUs during serving. Lower values are better for both metrics. In each experiment, we evaluate worst-case per-chunk latency under matched GPU operating cost, and GPU operating cost under matched worst-case per-chunk latency constraints.

\vspace{0.25em}
\noindent\textbf{Baseline implementations.} Since no prior work specifically focuses on serving infrastructure design for streaming video generation tasks, we compare \sys against three representative generic serving strategies adapted to our setting: (\textbf{\underline{i}}) \textbf{\textsc{TurboServe}\textsubscript{base}:} basic runtime in~\S\ref{subsec:basic-runtime}. Newly activated sessions are assigned to GPUs in round-robin order, while each GPU executes sessions in a first-come-first-served (FCFS) manner without migration or autoscaling. (\textbf{\underline{ii}}) \textbf{\textsc{TurboServe}\textsubscript{base}+LAG} (Load-Aware Greedy): newly activated sessions are assigned to the currently least-loaded GPU according to runtime GPU utilization, providing a stronger load-balancing baseline. (\textbf{\underline{iii}}) \textbf{\textsc{TurboServe}\textsubscript{base}+MAG} (\seqsplit{Memory-Aware} Greedy): newly activated sessions are assigned to the GPU with the lowest memory utilization, aiming to balance memory pressure and avoid GPU memory exhaustion under long-lived session workloads.
In addition, we further ablate the effectiveness of individual \sys mechanisms in \S\ref{subsec:ablation-exp}, including (\textbf{\underline{i}}) \textbf{\sys (w/o autoscaling)} and (\textbf{\underline{ii}}) \textbf{\sys (w/o migration)}.

\subsection{End-to-end Experimental Results}
\label{subsec:e2e-exp}

\noindent\textbf{End-to-end latency performance.} We first compare \sys against the baseline systems in terms of worst-case per-chunk latency under matched GPU operating cost, across different model sizes, workload traces, and cluster configurations. As shown in~\autoref{fig:e2e} (rows 1-2), \sys consistently outperforms all baseline systems, reducing worst-case per-chunk latency by 37.5\% on average and up to 51.6\% across all baseline comparisons. For example, on the 1.3B model with Trace~1 on Cluster~2, \sys reduces latency by 20.5\% and 26.6\% compared with the \textsc{TurboServe}\textsubscript{base}+LAG and \textsc{TurboServe}\textsubscript{base}+MAG baselines, respectively, and by 28.2\% compared with \textsc{TurboServe}\textsubscript{base}. Similar trends are observed across all evaluated traces and model sizes. These improvements come from \sys's ability to continuously rebalance sessions according to runtime load conditions, preventing bottleneck GPUs from emerging as session activity evolves.

\begin{figure*}[t!]
    \centering
    \includegraphics[width=\linewidth]{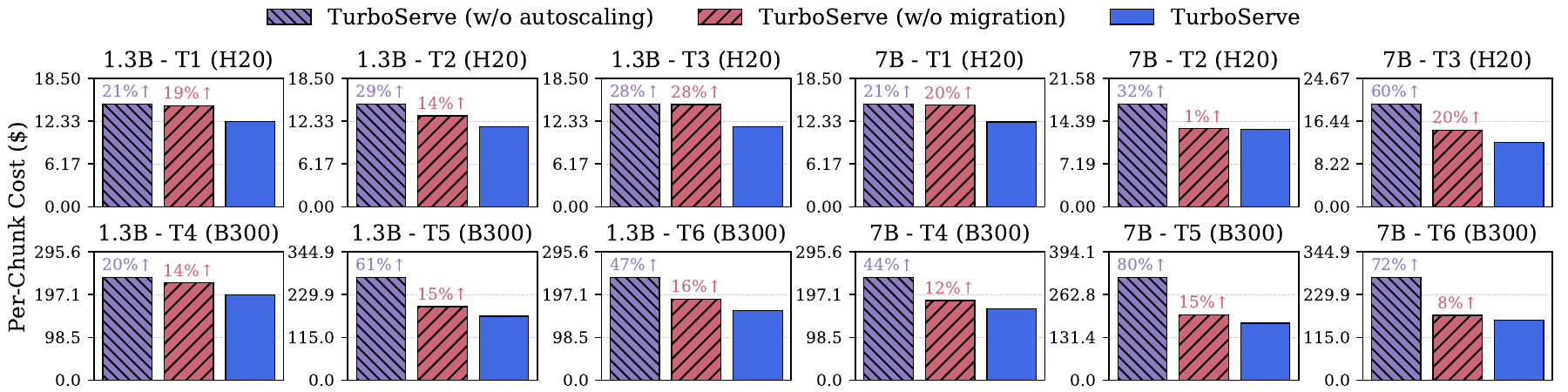}
    \caption{Cost efficiency ablation of \sys mechanisms, including session migration and GPU autoscaling, across different model sizes and workload traces under matched latency constraints.}
    \label{fig:ablation}
\end{figure*}

\vspace{0.25em}
\noindent\textbf{End-to-end cost efficiency.} We next compare \sys against the baseline systems in terms of GPU operating cost under matched worst-case per-chunk latency constraints, across different model sizes, workload traces, and cluster configurations. As shown in~\autoref{fig:e2e} (rows 3-4), \sys consistently achieves the lowest serving cost among all evaluated methods, reducing GPU operating cost by 37.2\% on average and up to 49.0\% across all baseline comparisons. For example, on the 1.3B model with Trace~1 on Cluster~2, \sys reduces GPU cost by 38.3\%, 35.9\%, and 16.7\% compared with \textsc{TurboServe}\textsubscript{base}, \textsc{TurboServe}\textsubscript{base}+LAG, and \textsc{TurboServe}\textsubscript{base}+MAG, respectively, while maintaining the same worst-case per-chunk latency target. Similar trends are observed across all evaluated traces and model sizes. These improvements stem from \sys's ability to dynamically adjust the GPU budget according to workload demand while simultaneously rebalancing sessions across available resources. By avoiding both resource underutilization during low-demand periods and excessive latency during bursts, \sys achieves a more efficient latency-cost trade-off than static serving strategies.

\vspace{0.5em}
\begin{mdframed}[style=analysisbox]
\small
\noindent\textit{\underline{Analysis:}} The end-to-end results show that \sys remains effective across model sizes, traces, and GPU clusters. This suggests that production streaming serving needs both latency control and cost control: by jointly rebalancing sessions and adapting GPU capacity, \sys improves latency stability under fixed cost and reduces GPU cost under fixed latency targets.
\end{mdframed}
\vspace{0.25em}

\subsection{Mechanism Ablation Study}
\label{subsec:ablation-exp}

\noindent\textbf{Ablation of migration and autoscaling.} We next evaluate the contribution of \sys's two key mechanisms, session migration and GPU autoscaling, through ablation experiments across different model sizes, workload traces, and GPU types. As shown in~\autoref{fig:ablation}, removing either mechanism generally degrades cost efficiency compared with the full system. Disabling migration increases GPU operating cost by 15.0\% on average and up to 28.0\%, because the system can no longer effectively rebalance sessions across GPUs, resulting in persistent load imbalance under the same latency constraint. Disabling autoscaling increases GPU operating cost by 42.9\% on average and up to 80.4\%, by preventing the system from adapting the GPU budget to workload fluctuations and forcing resources to be provisioned more conservatively to maintain latency guarantees. Across all evaluated models and traces, the full \sys configuration achieves the best cost efficiency. 

\vspace{0.5em}
\begin{mdframed}[style=analysisbox]
\small
\noindent\textit{\underline{Analysis:}} In production streaming workloads, migration and autoscaling address different system bottlenecks. Migration prevents persistent GPU imbalance, while autoscaling adapts capacity to workload variation; together, they enable \sys to consistently achieve the best cost efficiency across different models, workload traces, and clusters.
\end{mdframed}
\vspace{0.25em}

\subsection{Effectiveness of Closed-Loop Scheduling}
\label{subsec:scheduling-exp}

\begin{figure}[t!]
    \centering
    \includegraphics[width=0.5\linewidth]{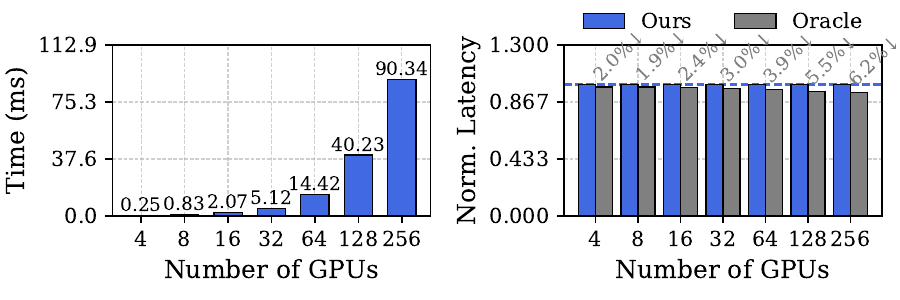}
    \caption{\textbf{\underline{Left:}} Scheduling efficiency of the migration-aware min-max rebalancing. \textbf{\underline{Right:}} Scheduling effectiveness of the migration-aware min-max rebalancing (vs. oracle).}
    \label{fig:efficiency}
    \label{fig:scheduling-effectivess}
\end{figure}

\noindent \textbf{Scheduling efficiency of the rebalancing algorithm.} As shown in~\autoref{fig:efficiency} (Left), we evaluate the scheduling efficiency of our migration-aware min-max rebalancing algorithm across a range of GPU scales (from 4 to 256 GPUs). The results show that our algorithm is fast enough for online video generation tasks and scales efficiently to large clusters. On clusters with up to 64 GPUs, scheduling completes within 15~ms, less than 2\% of the per-chunk generation time (on the order of hundreds of milliseconds). Even on larger clusters with 256 GPUs, scheduling completes within 0.1~s. In the rare cases where scheduling time becomes a bottleneck, we can partition the cluster into sub-groups (e.g., 64 GPUs each) and schedule each sub-group independently using additional CPU threads.



\vspace{0.25em}
\noindent \textbf{Scheduling effectiveness of the rebalancing algorithm.} To evaluate the effectiveness of our scheduling approach, we compare our migration-aware min-max rebalancing algorithm against an oracle that performs an exhaustive search for the best session placement across different cluster sizes. The oracle traverses all possible session placements across all GPUs to obtain the optimal solution, but at a much higher computational cost. We run each method on Trace 1 for 80 times and report the mean gap. As shown in~\autoref{fig:scheduling-effectivess} (Right), \sys closes the gap to the oracle to 3.6\% on average (6.5\% maximum), while reducing the scheduling time by more than 10$\times$ on average. 


\vspace{0.5em}
\begin{mdframed}[style=analysisbox]
\small
\noindent\textit{\underline{Analysis:}} In real-world streaming serving, scheduling latency is part of the serving critical path. An overly expensive placement optimizer may produce slightly better placements, but its runtime overhead can offset the benefit and hurt end-to-end serving efficiency. The results show that \sys strikes a practical system balance: it achieves near-oracle placement quality while keeping scheduling overhead low enough for online deployment.
\end{mdframed}
\vspace{0.25em}

\begin{table}[htbp]
\centering
\caption{Autoscaling cost (USD) of \sys compared with the offline oracle on Cluster 2 across three traces. Lower is better.}
\label{tab:autoscaling-effectiveness}
\resizebox{0.6\columnwidth}{!}{%
\begin{tabular}{l|c|c|c}
\hline
\textbf{Method} & \textbf{Trace 1} & \textbf{Trace 2} & \textbf{Trace 3} \\
\hline
\textbf{Oracle} & 188.03 \$ & 158.55 \$ & 160.74 \$ \\
\hline
\textbf{\sys}   & 196.87 \$ ($\uparrow$4.7\%) & 171.71 \$ ($\uparrow$8.3\%) & 152.36 \$ ($\uparrow$5.5\%) \\
\hline
\end{tabular}%
}
\end{table}

\vspace{0.25em}
\noindent \textbf{Scheduling effectiveness of the autoscaling mechanism.} To evaluate the effectiveness of our autoscaling mechanism, we compare its resource cost against an offline lower-bound oracle. Given the full trace in advance, including all future session arrivals, resumes, suspends, and departures, the oracle computes the minimum number of GPUs needed at each 1-minute slot to satisfy the target GPU utilization, then solves a dynamic program over GPU budgets to obtain the cost-optimal scaling schedule. We run both methods on three production traces on Cluster 2 and report the resulting cost in~\autoref{tab:autoscaling-effectiveness}. As shown, \sys stays within 6.1\% of the oracle cost on average (8.3\% maximum), despite operating online without any knowledge of future requests. 


\vspace{0.5em}
\begin{mdframed}[style=analysisbox]
\small
\noindent\textit{\underline{Analysis:}} In production streaming workloads, future demand is unavailable and can change rapidly. The small gap to the offline oracle shows that \sys captures most of the cost-saving opportunity using only online load feedback, making its autoscaling policy practical for real-world deployment.
\end{mdframed}
\vspace{0.25em}

\subsection{Runtime Breakdown and Overhead Analysis}
\label{subsec:breakdown-exp}

\begin{table}[htbp]
  \centering
  \small
  \caption{Twelve-minute snapshot of scheduling decisions for \sys in the Trace~1 experiment on Cluster~1. The autoscaling-decision column reports the actual GPU-budget changes in each two-minute window. The migration-decision column reports representative session movements, where $s$ denotes a session ID and $g$ denotes a GPU ID.}
  \label{tab:ours-scheduling-snapshot}
  \resizebox{0.6\columnwidth}{!}{%
  \begin{tabular}{l | l | l}
  \hline
  \textbf{Time Window} & \textbf{Autoscaling Decisions} & \textbf{Migration Decisions} \\
  \hline
  $(0, 2]$ min &
  $8{\rightarrow}1{\rightarrow}2{\rightarrow}3{\rightarrow}4{\rightarrow}5{\rightarrow}6$ &
  $s1{:}g1{\rightarrow}g0$, $s2{:}g1{\rightarrow}g0$, $\ldots$ $(+32)$ \\
  \hline
  $(2, 4]$ min &
  $6{\rightarrow}7{\rightarrow}8{\rightarrow}9{\rightarrow}10$ &
  $s49{:}g1{\rightarrow}g2$, $s44{:}g1{\rightarrow}g0$, $\ldots$ $(+21)$ \\
  \hline
  $(4, 6]$ min &
  $10{\rightarrow}11$ &
  $s18{:}g2{\rightarrow}g5$, $s52{:}g6{\rightarrow}g8$, $\ldots$ $(+22)$ \\
  \hline
  $(6, 8]$ min &
  $11{\rightarrow}10{\rightarrow}9{\rightarrow}7{\rightarrow}8$ &
  $s111{:}g7{\rightarrow}g5$, $s16{:}g2{\rightarrow}g7$, $\ldots$ $(+27)$ \\
  \hline
  $(8, 10]$ min &
  $8{\rightarrow}6{\rightarrow}7{\rightarrow}6{\rightarrow}4$ &
  $s111{:}g5{\rightarrow}g4$, $s45{:}g5{\rightarrow}g2$, $\ldots$ $(+23)$ \\
  \hline
  $(10, 12]$ min &
  $4{\rightarrow}5{\rightarrow}7{\rightarrow}8{\rightarrow}7{\rightarrow}6{\rightarrow}7$ &
  $s102{:}g1{\rightarrow}g3$, $s102{:}g3{\rightarrow}g2$, $\ldots$ $(+68)$ \\
  \hline
  \end{tabular}%
  }
  \end{table}

\noindent\textbf{Runtime scheduling breakdown.}
\autoref{tab:ours-scheduling-snapshot} presents a representative 12-minute snapshot of \sys's scheduling decisions on Trace~1 running on Cluster~1. The autoscaling decisions show that \sys continuously adjusts the active GPU budget according to workload demand. For example, the budget first contracts from 8 GPUs and then gradually expands during the initial window, continues scaling out as the workload increases in $(2,6]$ minutes, and later scales in during lower-demand periods. The migration decisions demonstrate how the placement controller reacts to these budget changes and runtime load imbalance by moving sessions across GPUs. Most windows involve tens of session movements, and the final window contains more movements because session placement becomes more imbalanced under the current GPU budget. 

\vspace{0.5em}
\begin{mdframed}[style=analysisbox]
\small
\noindent\textit{\underline{Analysis:}} In real-world streaming workloads, GPU demand and session-placement imbalance evolve at different timescales. The scheduling snapshot shows that autoscaling gradually adjusts cluster capacity according to workload demand, while migration frequently refines session placement to correct imbalance. This highlights a key system requirement: online deployment needs both coarse-grained resource adaptation and fine-grained session rebalancing.
\end{mdframed}

\begin{table}[htbp]
  \centering
  \small
  \caption{Session migration overhead across different cluster and model configurations on Trace~1. Per-chunk latency and migration overhead report the average chunk-generation latency and average GPU-GPU session migration latency, respectively.}
  \label{tab:migration-overhead}
  \resizebox{0.6\columnwidth}{!}{%
  \begin{tabular}{l | c | c | c | c}
  \hline
  \textbf{Metric} &
  \textbf{H20 (1.3B)} &
  \textbf{H20 (7B)} &
  \textbf{B300 (1.3B)} &
  \textbf{B300 (7B)} \\
  \hline
  Per-chunk Latency &
  1054 ms &
  1201 ms &
  917 ms &
  1181 ms \\
  \hline
  Migration Overhead &
  23 ms (2\%) &
  24 ms (2\%) &
  24 ms (3\%) &
  30 ms (3\%) \\
  \hline
  \end{tabular}%
  }
\end{table}

\noindent\textbf{Migration overhead analysis.} \autoref{tab:migration-overhead} reports the average session-migration overhead across different cluster and model configurations. We measure the end-to-end migration latency during online GPU-GPU rebalancing, including buffer allocation/release, NCCL-based state transfer, and session ownership updates. Across all settings, the migration overhead remains a small fraction of the per-chunk generation latency, ranging from 23-30 ms (2\%-3\% of per-chunk latency). Larger models incur higher absolute migration overhead due to larger per-session states, but the overhead remains a small and stable fraction of per-chunk latency across both H20 and B300 deployments.

\vspace{0.5em}
\begin{mdframed}[style=analysisbox]
\small
\noindent\textit{\underline{Analysis:}} In modern production clusters, GPU-GPU migration overhead is small relative to per-chunk generation latency, while the latency penalty from persistent GPU imbalance can be substantial. This makes session migration practical as a regular system control action, as long as migrations are triggered only when the expected latency reduction outweighs the transfer cost.
\end{mdframed}
\vspace{0.25em}

\vspace{-0.5em}
\section{Conclusion}
\label{sec:conclusion}

This paper presents \sys, the \textit{first} serving system designed specifically for streaming video generation workloads in multi-session, multi-GPU environments. \sys addresses the challenges of heterogeneous session duration and time-varying user demand through a closed-loop scheduling framework that jointly coordinates migration-aware session placement and load-driven GPU autoscaling. It further supports efficient runtime execution through coalesced chunk processing and session state migration. Evaluations on real-world production traces from \companyname show that \sys reduces worst-case per-chunk latency by 37.5\% and total GPU operating cost by 37.2\% on average compared with baseline serving configurations. These results demonstrate the effectiveness of \sys in delivering cost-efficient and latency-stable serving for dynamic streaming video generation workloads.


\bibliographystyle{plainnat}
\bibliography{reference}

\clearpage

\appendix

\section{Volatility-to-Parameter Mapping}
\label{app:volatility-mapping}

The autoscaling controller adapts the control parameters $(\lambda(t), \hat{\rho}(t))$ in response to workload dynamics (\S\ref{subsubsec:autoscalingcontroller}). We measure workload variability through a single scalar metric, the \textit{volatility} $\sigma(t)$ defined below, and realize the adaptation through a volatility-to-parameter mapping $\mathcal{T}: \mathcal{V} \to \mathbb{R}_{>0} \times (0, 1]$ that associates each volatility level $v_{\ell} \in \mathcal{V} = \{v_{1}, \ldots, v_{L}\}$ (e.g., $L = 10$) with a profiled control parameters $(\lambda_{\ell}, \rho^{*}_{\ell})$. The mapping is constructed offline and queried online to update the active parameters.

\vspace{0.25em}
\noindent\textbf{Volatility metric.}
Let $a_{\tau}$ denote the number of newly activated sessions at event $\tau$. Given a sliding event window of size $W$ ending at event $t$, the workload volatility is defined as:
\begin{equation}
\sigma(t) = \mathrm{std}\bigl(a_{t-W+1}, \ldots, a_{t}\bigr).
\end{equation}
The volatility range observed across historical traces is partitioned into $L$ ordered intervals, with $v_{\ell}$ representing the $\ell$-th interval from low to high.

\vspace{0.25em}
\noindent\textbf{Offline profiling.}
We subsample representative segments from historical real-world workload traces and group them by volatility level. For each level $v_{\ell}$, we perform a grid search over candidate control parameters $(\lambda, \rho^{*})$, replay the corresponding trace segments through the scheduler, and select the parameters $(\lambda_{\ell}, \rho^{*}_{\ell})$ that minimizes GPU cost subject to the target worst-case per-chunk latency constraint. The resulting table $\{\mathcal{T}(v_{\ell}) = (\lambda_{\ell}, \rho^{*}_{\ell})\}_{\ell=1}^{L}$ is persisted and loaded by the controller at startup.

\vspace{0.25em}
\noindent\textbf{Online replacement workflow.}
At each event $t$, the controller updates the active control parameters via a four-step \textit{measure-quantize-look-up-replace} workflow:
\begin{enumerate}
    \item \textit{Measure.} Compute the current volatility $\sigma(t)$ over the sliding window of recent events.
    \item \textit{Quantize.} Map $\sigma(t)$ to the level $v_{\ell(t)} \in \mathcal{V}$ whose interval contains $\sigma(t)$.
    \item \textit{Look up.} Retrieve the profiled parameters $(\lambda_{\ell(t)}, \rho^{*}_{\ell(t)}) = \mathcal{T}(v_{\ell(t)})$.
    \item \textit{Replace.} Commit the update $(\lambda(t), \hat{\rho}(t)) \leftarrow (\lambda_{\ell(t)}, \rho^{*}_{\ell(t)})$, which takes effect in subsequent autoscaling decisions.
\end{enumerate}

\vspace{0.25em}
\noindent\textbf{Profiling case study.}
We illustrate the offline profiling and the resulting volatility-to-parameter mapping on a representative trace family consisting of $L = 10$ segments of monotonically increasing volatility, generated by progressively scaling the burst magnitude of a base activation pattern. \autoref{tab:case-study-workload} summarizes the workload characteristics of each segment, and \autoref{tab:case-study-mapping} reports the profiled control parameters together with the cost and latency observed during replay under the latency constraint $L_{\text{SLO}} = 670\,\text{ms}$. Three observations stand out from the profiled mapping:

\begin{table}[t]
\centering
\small
\caption{Workload characteristics of the ten profiling segments. \textit{Volatility} is the standard deviation of newly activated session counts across $5\,\text{s}$ bins; \textit{arrivals} is the total number of new sessions over the segment; \textit{peak active} is the maximum number of concurrent active sessions.}
\label{tab:case-study-workload}
\resizebox{0.4\columnwidth}{!}{%
\begin{tabular}{r | r | r | r}
\hline
\textbf{Level} & \textbf{Volatility} & \textbf{Arrivals} & \textbf{Peak active} \\
\hline
1  & 0.86 &  7 & 23 \\
\hline
2  & 1.32 & 11 & 27 \\
\hline
3  & 1.92 & 15 & 31 \\
\hline
4  & 2.66 & 19 & 35 \\
\hline
5  & 3.15 & 23 & 39 \\
\hline
6  & 3.77 & 27 & 43 \\
\hline
7  & 4.39 & 31 & 47 \\
\hline
8  & 5.14 & 35 & 51 \\
\hline
9  & 5.51 & 39 & 55 \\
\hline
10 & 6.38 & 43 & 59 \\
\hline
\end{tabular}
}
\end{table}

\begin{table}[t]
\centering
\small
\caption{Profiled control parameters and replay results per volatility level under the latency constraint $L_{\text{SLO}} = 670\,\text{ms}$. \textit{Valid} indicates whether the configuration satisfies the latency constraint; \textit{pass rate} is the fraction of generated chunks whose per-chunk latency falls within $L_{\text{SLO}}$; \textit{avg.\ cost} is the average GPU operating cost observed during replay.}
\label{tab:case-study-mapping}
\resizebox{0.6\columnwidth}{!}{%
\begin{tabular}{r | r | r | r | c | r | r}
\hline
\textbf{Level} & \textbf{Volatility} & \textbf{$\lambda$} & \textbf{$\rho^{*}$} & \textbf{Valid} & \textbf{Pass rate} & \textbf{Avg.\ cost} \\
\hline
1  & 0.86 & 0.2 & 0.80 & \checkmark & 100\% & 338.66 \\
\hline
2  & 1.32 & 0.2 & 0.80 & \checkmark & 100\% & 369.35 \\
\hline
3  & 1.92 & 0.2 & 0.65 & \checkmark & 100\% & 498.73 \\
\hline
4  & 2.66 & 0.2 & 0.65 & \checkmark & 100\% & 526.14 \\
\hline
5  & 3.15 & 0.2 & 0.65 & \checkmark & 100\% & 547.34 \\
\hline
6  & 3.77 & 0.2 & 0.50 & \checkmark & 100\% & 700.77 \\
\hline
7  & 4.39 & 0.2 & 0.50 & \checkmark & 100\% & 702.44 \\
\hline
8  & 5.14 & 0.2 & 0.50 & \checkmark & 100\% & 705.05 \\
\hline
9  & 5.51 & 0.2 & 0.25 & \checkmark & 100\% & 770.44 \\
\hline
10 & 6.38 & 0.2 & 0.25 & \checkmark & 100\% & 769.46 \\
\hline
\end{tabular}
}
\end{table}

\vspace{0.25em}
\noindent\textit{\underline{O1:} Monotonic reduction of $\rho^{*}$ with volatility.}
The target utilization $\rho^{*}_{\ell}$ decreases monotonically as the volatility level increases, dropping from $0.80$ at level 1 to $0.25$ at level 10. This directly realizes the design intuition stated in \S\ref{subsubsec:autoscalingcontroller}: at low volatility the controller can pack sessions densely without risking latency violations, while at high volatility it must reserve substantial per-GPU headroom to absorb activation bursts. Notably, the profiled $\rho^{*}_{\ell}$ values fall into four discrete bands ($0.80, 0.65, 0.50, 0.25$), each spanning two to three adjacent volatility levels, suggesting that the cost-latency frontier exhibits a piecewise-flat structure rather than continuously varying with volatility.

\vspace{0.25em}
\noindent\textit{\underline{O2:} Stable $\lambda$ across the entire range.}
The latency weight $\lambda_{\ell}$ remains fixed at $0.2$ across all ten levels, indicating that for this trace family the cost-latency trade-off is governed primarily by $\rho^{*}$, with $\lambda$ playing a secondary role. Conceptually, $\rho^{*}$ adapts the system's safety margin in capacity space, while $\lambda$ would primarily activate as a second axis under workloads where headroom adjustment via $\rho^{*}$ alone proves insufficient (e.g., when $\rho^{*}$ saturates at its lower bound and further headroom can only be obtained by penalizing latency more aggressively in the objective).

\vspace{0.25em}
\noindent\textit{\underline{O3:} Monotonic increase of cost with volatility.}
The average GPU cost increases monotonically as the volatility level rises, from $338.66$ at level 1 to $770.44$ at level 9, roughly $2.3\times$. This growth mirrors the reduction in $\rho^{*}$ documented in O1: as $\rho^{*}_{\ell}$ shrinks, each GPU carries fewer concurrent sessions, forcing the controller to provision additional GPUs to serve the same workload. The mapping therefore makes explicit the cost of accommodating burstiness: each volatility level pays exactly the additional GPU cost required to handle its bursts. A non-adaptive controller using a fixed $\rho^{*}$ would either incur level-9 cost at level-1 workloads (over-provisioning under stable load) or violate the latency constraint at high volatility (under-provisioning under bursts), whereas the volatility-keyed mapping recovers the cost-minimal feasible configuration at each level.

\vspace{0.25em}
\noindent \textit{Results analysis.} All ten levels achieve a 100\% pass rate under $L_{\text{SLO}} = 670\,\text{ms}$, confirming that the constrained grid search consistently identifies feasible $(\lambda, \rho^{*})$ configurations across the entire volatility range. The mapping is therefore simultaneously \textit{safe} (no per-chunk latency violations are observed), and \textit{cost-aware} (each level pays only the cost required to absorb its own characteristic burstiness).

\vspace{0.25em}
\noindent\textbf{Comparison with an offline oracle.}
We further evaluate how close the online mapping is to an offline trace-tuned oracle. The goal is to isolate the gap caused by online, adaptive control-parameter selection, rather than differences in scheduling primitives, so both methods use the same migration and autoscaling mechanisms, the same maximum GPU budget, and the same scale-out delay and cost model.

\vspace{0.25em}
\noindent\emph{Oracle.}
The oracle has access to the realized workload trace in each future control window. For every $30$\,s window, it searches over $\rho^{*} \in \{0.50, 0.55, \ldots, 0.95\}$ and selects the cost-minimizing value that still satisfies the latency constraint, knowing the future arrivals exactly. This baseline is not deployable in practice, but provides an upper bound for evaluating the quality of the online mapping.

\vspace{0.25em}
\noindent\emph{Ours.}
The online mapping does not observe the future trace. During serving, it computes the volatility from a recent sliding window, maps it to a profiled level, and applies the corresponding control parameters. In this experiment, $\lambda$ is fixed at $0.20$ and adaptation is performed through $\rho^{*}$ alone.

\vspace{0.25em}
\noindent\emph{Workload trace.}
We construct an unseen $5$-minute workload comprising ten $30$\,s windows that alternate between low, medium, and high pressure. Each GPU hosts at most five concurrent sessions, and the maximum GPU budget is $16$. \autoref{tab:oracle-workload} summarizes the per-window load characteristics.

\begin{table}[t]
\centering
\small
\caption{Unseen fluctuating workload used for the oracle comparison. \textit{Avg.}/\textit{max active sessions} characterize the per-window load.}
\label{tab:oracle-workload}
\resizebox{0.9\columnwidth}{!}{%
\begin{tabular}{l | r | r | r | r | r | r | r | r | r | r}
\hline
\textbf{Window} & \textbf{w1} & \textbf{w2} & \textbf{w3} & \textbf{w4} & \textbf{w5} & \textbf{w6} & \textbf{w7} & \textbf{w8} & \textbf{w9} & \textbf{w10} \\
\hline
Avg.\ active sessions & 32.00 & 17.17 & 7.67 & 23.47 & 51.23 & 72.43 & 12.43 & 56.90 & 22.30 & 53.17 \\
\hline
Max active sessions & 38 & 18 & 8 & 28 & 58 & 77 & 13 & 68 & 23 & 73 \\
\hline
\end{tabular}%
}
\end{table}

\vspace{0.25em}
\noindent\emph{Selected control parameters.}
\autoref{tab:oracle-rho} reports the target utilization selected by the oracle and by our online mapping. With full knowledge of future demand, the oracle picks the per-window cost-minimizing $\rho^{*}$, packing densely in low-pressure windows ($\rho^{*}$ up to $0.95$ in w3) and lowering $\rho^{*}$ to $0.55$ during high-pressure windows. Our method, observing only recent history, follows a qualitatively similar pattern in the early windows but settles into a steady $\rho^{*} = 0.60$ once volatility has risen, retaining this conservative setting until the volatility estimate decays. Overall, the oracle tends toward sharp, per-window-optimal swings (range $0.55$ to $0.95$), while our method tends toward smoother, more conservative selections (range $0.60$ to $0.88$), trading per-window optimality for stability against unforeseen bursts.

\begin{table}[t]
\centering
\small
\caption{Per-window target utilization $\rho^{*}$.}
\label{tab:oracle-rho}
\resizebox{0.8\columnwidth}{!}{%
\begin{tabular}{l | r | r | r | r | r | r | r | r | r | r}
\hline
\textbf{Window} & \textbf{w1} & \textbf{w2} & \textbf{w3} & \textbf{w4} & \textbf{w5} & \textbf{w6} & \textbf{w7} & \textbf{w8} & \textbf{w9} & \textbf{w10} \\
\hline
Offline oracle $\rho^{*}$ & 0.60 & 0.88 & 0.95 & 0.72 & 0.55 & 0.55 & 0.92 & 0.55 & 0.80 & 0.55 \\
\hline
Ours $\rho^{*}$ & 0.66 & 0.72 & 0.88 & 0.78 & 0.60 & 0.60 & 0.60 & 0.60 & 0.60 & 0.60 \\
\hline
\end{tabular}%
}
\end{table}

\begin{table}[t]
\centering
\small
\caption{End-to-end comparison between the online mapping and the offline oracle. \textit{Cost} is the total GPU operating cost; \textit{avg.}/\textit{max lat.} are the per-chunk latency statistics across the entire trace.}
\label{tab:oracle-results}
\resizebox{0.45\columnwidth}{!}{%
\begin{tabular}{l | r | r | r}
\hline
\textbf{Method} & \textbf{Cost} & \textbf{Avg.\ lat.} & \textbf{Max lat.} \\
\hline
Offline oracle & 3695.00 & 410.47\,ms & 668.05\,ms \\
\hline
Ours & 3721.93 & 421.57\,ms & 667.46\,ms \\
\hline
\end{tabular}%
}
\end{table}

\begin{table}[t]
\centering
\small
\caption{Window-end GPU budgets selected by each method.}
\label{tab:oracle-gpu}
\resizebox{0.7\columnwidth}{!}{%
\begin{tabular}{l | r | r | r | r | r | r | r | r | r | r}
\hline
\textbf{Window} & \textbf{w1} & \textbf{w2} & \textbf{w3} & \textbf{w4} & \textbf{w5} & \textbf{w6} & \textbf{w7} & \textbf{w8} & \textbf{w9} & \textbf{w10} \\
\hline
Offline oracle & 13 & 5 & 2 & 8 & 16 & 16 & 3 & 16 & 6 & 16 \\
\hline
Ours & 11 & 8 & 8 & 8 & 16 & 16 & 8 & 16 & 8 & 15 \\
\hline
\end{tabular}%
}
\end{table}

\vspace{0.25em}
\noindent\emph{Experimental results.}
\autoref{tab:oracle-results} reports the headline simulation outcomes, and \autoref{tab:oracle-gpu} shows the window-end GPU budgets. Both methods keep the served-chunk maximum latency at or below $668.05$\,ms, within the $670$\,ms target, with comparable average latencies ($410.47$\,ms for the oracle and $421.57$\,ms for our method). The oracle uses fewer GPUs in low-pressure windows where it can safely raise $\rho^{*}$ (e.g., w2, w3, w7, w9), while our method holds additional headroom in those windows because the recent volatility estimate has not yet decayed. On total operating cost, the online mapping ($3721.93$) comes within $0.73\%$ of the offline oracle ($3695.00$).

\vspace{0.25em}
\noindent\emph{Results analysis.}
This near-equality is striking given the substantial information asymmetry between the two methods: the oracle has access to every future arrival in every window and selects $\rho^{*}$ per-window optimally (ranging from $0.55$ to $0.95$), whereas our method observes only a recent sliding window of activations and chooses $\rho^{*}$ from the volatility mapping (ranging from $0.60$ to $0.88$). The result demonstrates that the volatility-based mapping is robust across diverse workload regimes: it handles low-pressure intervals, sustained bursts, and abrupt transitions alike without exceeding the latency target, while incurring almost the same total cost as the oracle. The mapping therefore captures essentially all of the cost-relevant structure that future knowledge could exploit, and remains effective under the inherent delay between observing rising volatility and provisioning additional GPUs. Since the oracle requires future knowledge and is not deployable in practice, it represents an unrealizable upper bound, and the online mapping effectively closes the gap to it under realistic causality constraints. We also conduct the same comparison on two additional traces subsampled from real-world workloads, where the cost gap to the oracle is $2.25\%$ and $3.99\%$, indicating that our volatility-based mapping generalizes beyond this single trace.

\section{Workload Statistics}
\label{appendix:workload-statistics}

\begin{table}[t]
\centering
\caption{Workload trace characteristics. \textit{Arrivals} is the number of new jobs entering the system in each span; \textit{Departures} is the number of jobs completing in each span; \textit{Avg.\ active} is the mean number of concurrently active jobs over the span.}
\label{tab:trace_stats}
\resizebox{0.5\linewidth}{!}{%
\begin{tabular}{l | c | c | c | c | c}
\hline
Time (min) & $[0,2)$ & $[2,4)$ & $[4,6)$ & $[6,8)$ & $[8,10]$ \\
\hline
Arrivals      & 31     & 47       & 30       & 48       & 44        \\
\hline
Departures    & 17     & 24       & 26       & 39       & 49        \\
\hline
Avg.\ active  & 10.36  & 20.91    & 19.30    & 29.62    & 33.49     \\
\hline
\end{tabular}%
}
\end{table}

\autoref{tab:trace_stats} demonstrates the workload trace characteristics for characterization and motivation.

\autoref{tab:trace_stats_full} reports the detailed characteristics of the evaluated traces. Traces T1--T3 are evaluated on the H20 cluster (Cluster~1) and traces T4--T6 on the B300 cluster (Cluster~2)

\begin{table}[t]
\centering
\caption{Workload trace characteristics. \textit{Arrivals} is the number of new jobs entering the system in each span; \textit{Departures} is the number of jobs completing in each span; \textit{Avg.\ active} is the mean number of concurrently active jobs over the span.}
\label{tab:trace_stats_full}
\resizebox{0.6\linewidth}{!}{%
\begin{tabular}{c | l | c | c | c | c | c | c}
\hline
Trace & Metric & $[0,1)$ & $[1,2)$ & $[2,3)$ & $[3,4)$ & $[4,5)$ & Total \\
\hline
\multirow{3}{*}{T1}
 & Arrivals     & 122   & 130   & 66    & 22    & 18    & 358  \\
 & Departures   & 44    & 86    & 88    & 64    & 38    & 320  \\
 & Avg.\ active & 28.0  & 56.2  & 57.4  & 37.4  & 23.2  & --   \\
\hline
\multirow{3}{*}{T2}
 & Arrivals     & 218   & 214   & 248   & 192   & 204   & 1076 \\
 & Departures   & 60    & 124   & 204   & 196   & 204   & 788  \\
 & Avg.\ active & 61.0  & 118.6 & 147.6 & 154.2 & 149.4 & --   \\
\hline
\multirow{3}{*}{T3}
 & Arrivals     & 74    & 148   & 156   & 264   & 156   & 798  \\
 & Departures   & 10    & 64    & 126   & 148   & 214   & 562  \\
 & Avg.\ active & 13.2  & 49.4  & 112.8 & 121.4 & 148.4 & --   \\
\hline
\multirow{3}{*}{T4}
 & Arrivals     & 500   & 428   & 308   & 88    & 80    & 1404 \\
 & Departures   & 158   & 266   & 388   & 236   & 152   & 1200 \\
 & Avg.\ active & 118.4 & 219.2 & 268.0 & 162.8 & 101.8 & --   \\
\hline
\multirow{3}{*}{T5}
 & Arrivals     & 874   & 862   & 998   & 762   & 814   & 4310 \\
 & Departures   & 238   & 504   & 810   & 784   & 814   & 3150 \\
 & Avg.\ active & 245.8 & 475.2 & 589.2 & 616.8 & 598.4 & --   \\
\hline
\multirow{3}{*}{T6}
 & Arrivals     & 296   & 590   & 626   & 1062  & 618   & 3192 \\
 & Departures   & 36    & 260   & 502   & 608   & 846   & 2252 \\
 & Avg.\ active & 54.4  & 198.4 & 451.4 & 487.2 & 592.0 & --   \\
\hline
\end{tabular}%
}
\end{table}

\end{document}